\documentclass[aps, prd, preprint, amsfonts, floatfix]{revtex4}

\usepackage{graphicx}
\usepackage{amsmath,amsfonts}
\usepackage{epsfig,color}

\begin{document}
\newcommand{\be}{\begin{eqnarray}}
\newcommand{\ee}{\end{eqnarray}}
\newcommand\del{\partial}
\newcommand\nn{\nonumber}
\newcommand{\Tr}{{\rm Tr}}
\newcommand{\Str}{{\rm Trg}}
\newcommand{\mat}{\left ( \begin{array}{cc}}
\newcommand{\emat}{\end{array} \right )}
\newcommand{\vect}{\left ( \begin{array}{c}}
\newcommand{\evect}{\end{array} \right )}
\newcommand{\tr}{{\rm Tr}}
\newcommand{\hm}{\hat m}
\newcommand{\ha}{\hat a}
\newcommand{\hz}{\hat z}
\newcommand{\hze}{\hat \zeta}
\newcommand{\hx}{\hat x}
\newcommand{\hy}{\hat y}
\newcommand{\tm}{\tilde{m}}
\newcommand{\ta}{\tilde{a}}
\newcommand{\U}{\rm U}
\newcommand{\diag}{{\rm diag}}
\newcommand{\tz}{\tilde{z}}
\newcommand{\tx}{\tilde{x}}
\definecolor{red}{rgb}{1.00, 0.00, 0.00}
\newcommand{\rd}{\color{red}}
\definecolor{blue}{rgb}{0.00, 0.00, 1.00}
\definecolor{green}{rgb}{0.10, 1.00, .10}
\newcommand{\blu}{\color{blue}}
\newcommand{\green}{\color{green}}



\title{The Realization of the Sharpe-Singleton Scenario}
\author{M. Kieburg}
\affiliation{Department of Physics and Astronomy, SUNY, Stony Brook,
 New York 11794, USA}
\author{K. Splittorff}
\affiliation{Discovery Center, The Niels Bohr Institute, University of Copenhagen, 
Blegdamsvej 17, DK-2100, Copenhagen {\O}, Denmark} 
\author{J.J.M. Verbaarschot}
\affiliation{Department of Physics and Astronomy, SUNY${  }$, Stony Brook,
 New York 11794, USA}

\date   {\today}
\begin  {abstract}
The microscopic spectral density of the Wilson Dirac operator 
for two flavor lattice QCD is analyzed. The computation 
includes the leading order $a^2$ corrections of the chiral Lagrangian
in the microscopic limit. 
The result is used to demonstrate how the Sharpe-Singleton first 
order scenario is realized in terms of the eigenvalues of the Wilson Dirac 
operator. We show that the Sharpe-Singleton scenario only takes place 
in the theory with dynamical fermions whereas the Aoki phase 
can be realized in the 
quenched as well as the unquenched theory. Moreover, we give  constraints 
imposed by $\gamma_5$-Hermiticity on the additional low energy constants 
of Wilson chiral perturbation theory.

\end{abstract}
\maketitle

\section{Introduction}

In the deep chiral limit, with almost massless quarks, lattice 
QCD with Wilson fermions has a highly nontrivial phase structure.
As in continuum QCD, it is the deep chiral limit which reveals the 
spontaneous breaking of chiral symmetry on the lattice. In addition, 
the interplay between the continuum and the chiral limit in 
lattice QCD with Wilson fermions leads to new phase structures 
known as the Aoki phase \cite{Aokiclassic} and the 
Sharpe-Singleton scenario \cite{SharpeSingleton}. These phases have 
no direct analogues in the continuum theory, and dominate if the 
chiral limit is performed prior to the continuum limit.
While this at first may seem like a highly undesirable artifact 
of Wilson fermions it can in fact be turned to our advantage: 
The Aoki phase is reached through a second order phase transition 
and at the boundary of this transition the pions are massless. 
This opens the possibility to study nonperturbative QCD at 
extremely small pion masses even at a nonzero lattice spacing.
On the contrary the Sharpe-Singleton scenario is a first order 
phase transition in which the pions are massive even in the 
chiral limit at nonzero lattice spacing. 

These phase structures of lattice QCD with Wilson fermions can 
be described within the framework of Wilson chiral perturbation 
theory \cite{SharpeSingleton,RS,BRS,Aoki-spec,GSS,Shindler,BNSep}. 
This low energy 
effective theory of lattice QCD with Wilson fermions describes 
 discretization effects by means of additional terms in 
the chiral Lagrangian (see \cite{Golterman,sharpe-nara} for 
reviews). Each of these new terms come with a new low energy 
constant. The sign and magnitude of these constants reflect whether 
lattice QCD with Wilson fermions will enter the Aoki phase or the 
Sharpe-Singleton scenario. Considerable progress, both analytically 
\cite{DSV,ADSVprd,HS1,HS2,SVlat2011,SVtwist} 
and numerically \cite{NS,arXiv:0709.4564,757044,arXiv:0911.5061,DWW,DHS,BBS}, 
has been made recently in the determination of these constants. However, 
a complete picture has not yet emerged.
For example, the observation that quenched lattice simulations 
consistently observe the Aoki phase 
\cite{Aoki:1992nb,Aoki:1990ap,Aoki:1989rw,Jansen:2005cg}, while in 
unquenched simulations
both the Aoki and the Sharpe-Singleton scenario \cite{Aoki:1995yf,Aoki:1997fm,Ilgenfritz:2003gw,CERN1,CERN2,CERN3,Aoki:2004iq,BBS,Farchioni:2004us,Farchioni:2004fs,Farchioni:2005tu,arXiv:0911.5061} has been observed, 
remains a puzzle.

The spontaneous breaking of chiral symmetry is tightly connected 
to the smallest eigenvalues of the Dirac operator \cite{BC,Giusti:2008vb}. 
Moreover, the Aoki phase manifests itself in the smallest eigenvalues 
of the Wilson Dirac operator \cite{Heller,DSV}. Here 
we show that the behavior of the smallest eigenvalues of the Wilson 
Dirac operator is also directly related to the Sharpe-Singleton 
scenario. In particular, we explain that in the Sharpe-Singleton scenario
the Wilson Dirac eigenvalues undergo  a collective macroscopic 
jump as the quark mass changes sign. Moreover, we 
show that this collective jump only occurs in the presence of dynamical 
fermions. The quenched theory has no analogue of this and hence 
the Sharpe-Singleton scenario is not possible in the 
quenched theory. This conclusion is verified by a direct computation 
of the microscopic quenched and unquenched chiral condensate.

In order to establish these results we explicitly derive 
the unquenched microscopic spectral density of the Wilson 
Dirac operator. This calculation makes use of both Wilson 
random matrix theory as well as Wilson chiral perturbation 
theory. By means of an underlying Pfaffian structure we 
uncover a compact factorized form of the exact unquenched 
microscopic eigenvalue density. This form makes it possible 
to understand the full dependence of the eigenvalue density 
on the low energy constants. We analyze this dependence in 
the mean field limit which can also be directly derived
from Wilson chiral perturbation theory. 

The mean field limit of the 
microscopic spectral density corresponds to the leading order result 
of Wilson chiral perturbation theory in the $p$-regime. This 
will allow us to close the circle by explaining the original 
$p$-regime results of Sharpe and Singleton in terms of 
the behavior of the Wilson Dirac eigenvalues. In particular,
we will explain how the nonzero minimal 
value of the pion mass in the Sharpe-Singleton scenario is 
connected to the collective jump of the Wilson Dirac eigenvalues.

The approach to the Wilson Dirac spectrum followed in this paper has been 
applied previously in Refs.~\cite{sharpe,DSV,ADSVprd,NS,ADSVNf1,SV-Nf2,KVZ,AN,SVlat2011,SVtwist,K} and results from these studies will be used.

The study of the smallest eigenvalues of the Wilson Dirac 
eigenvalues not only explains the way in which the Aoki phase 
and the Sharpe-Singleton scenarios are realized, it also 
gives direct information on the sign and magnitude of the 
low energy constants of Wilson chiral perturbation theory.
We will show that the spectral properties of the Wilson 
Dirac operator determine the sign of all three additional 
low energy constants of the leading order chiral Lagrangian
of Wilson chiral perturbation theory in the microscopic limit.

The results for the unquenched spectral density of the Wilson 
Dirac operator presented here also offer a direct way to measure 
the low energy constants of Wilson chiral perturbation theory by
matching the predictions against results from lattice QCD. The 
first quenched studies of this nature appeared recently 
\cite{DWW,DHS}.

This paper is organized as follows.
After a brief presentation of the properties of the Wilson Dirac 
operator in Section \ref{sec:DW} we recall the basics of Wilson 
chiral perturbation theory in section \ref{sec:WCPT}.
In section \ref{sec:constraints} we determine  constraints on 
the additional low energy parameters of Wilson chiral perturbation 
theory in terms of the spectral properties of the Wilson Dirac 
operator. The unquenched microscopic spectrum of the Wilson 
Dirac operator is analyzed in section \ref{sec:rho_c}. Finally, 
the realization of the Sharpe-Singleton scenario is the topic of 
section \ref{sec:SharpeSingleton}. Section \ref{sec:conc} contains 
our summary and conclusions. 
Wilson random matrix theory, the factorization properties of the spectral density and
the details of the mean field calculation 
are discussed in \ref{app:WRMT}, \ref{app:ZNf} and \ref{app:MF}, respectively.

\section{The Wilson Dirac operator}
\label{sec:DW}

Here we recall a few basic properties of the Wilson Dirac operator. 
The Wilson term in the lattice discretized covariant derivative
\be
D_W =\frac{1}{2}\gamma_\mu(\nabla_\mu+\nabla_\mu^*)
     -\frac{ar}{2}\nabla_\mu\nabla_\mu^*
\ee
breaks the anti-Hermiticity as well as the 
axial symmetry of the continuum Dirac operator. However, $D_W$ is 
$\gamma_5$-Hermitian 
\be
\gamma_5D_W\gamma_5=D_W^\dagger
\ee
and the product with $\gamma_5$, $D_5(m)\equiv\gamma_5(D_W+m)$ is 
therefore Hermitian.   

The eigenvalues, $z_k$, of $D_W$ consists of complex conjugated pairs as well 
as exactly real eigenvalues \cite{Itoh}. Only the real eigenmodes have 
nonzero chirality and  determine the index, $\nu$, of the Wilson 
Dirac operator 
\be
\label{defIndex}
\nu = \sum_{k} {\rm sign} (\langle k|\gamma_5|k\rangle).
\ee
Here $|k\rangle$ denotes the $k$'th eigenstate of $D_W$. 
The eigenvalues, $\lambda^5$, of $D_5(m)$ are unpaired when $a\neq0$.

In section \ref{sec:constraints} 
below we will use these properties to constrain the parameters of Wilson 
chiral perturbation theory.

\section{Wilson Chiral Perturbation Theory}
\label{sec:WCPT}

In the microscopic limit at nonzero lattice spacing where 
($m$ is the quark mass, $\zeta$ the axial quark mass, $z$ 
an eigenvalue of $D_W$, and $a$ is the lattice spacing)  
\be
 mV, \quad  \zeta V, \quad  z V \quad {\rm and} \quad a^2V 
\ee
are kept fixed as $V\to\infty$, the microscopic partition function of 
\cite{GL} extends to \cite{DSV}
\be
\label{Znu}
Z_{N_f}^\nu(m,\zeta;a) =   \int_{U(N_f)} \hspace{-1mm} d U \ {\det}^\nu U
~e^{S[U]}, 
\ee
where the action $S[U]$ for degenerate quark masses is given by 
\cite{SharpeSingleton,RS,BRS}
\be\label{lfull}
S & = & \frac{m}{2}\Sigma V{\rm Tr}(U+U^\dagger)+
\frac{\zeta}{2}\Sigma V{\rm Tr}(U-U^\dagger)\\
&&-a^2VW_6[{\rm Tr}\left(U+U^\dagger\right)]^2
     -a^2VW_7[{\rm Tr}\left(U-U^\dagger\right)]^2 
-a^2 V W_8{\rm Tr}(U^2+{U^\dagger}^2) .\nn
\ee
In addition to the chiral condensate, $\Sigma$, the action also 
contains the low energy constants $W_6$, $W_7$ and $W_8$ as parameters 
\footnote{Note that we use the convention
  of \cite{DSV,ADSVprd} for the low energy constants $W_6$, $W_7$ and
  $W_8$. In \cite{BRS} these constants are denoted by $-{W_6}'$,
  $-{W_7}'$ and $-{W_8}'$ respectively.}.

In order to lighten the notation we introduce the rescaled, dimensionless 
variables
\be
  \ha_i^2 = a^2 V W_i, \qquad  \hm = mV \Sigma, \qquad  \hz = zV \Sigma \quad {\rm and}   \qquad
 \hze = \zeta V \Sigma .
\ee

The generating functional for the eigenvalue density of $D_W$ in the 
complex plane is the graded extension of Eq. (\ref{Znu}).  
Because of the non-Hermiticity of $D_W$, the graded extension 
\be
\label{ZnuNf+2|2}
Z_{N_f+2|2}^\nu(\hz,\hz^*,\hz',\hz'^*,\hm;\ha_i)
\ee
requires an extra pair of conjugate quarks with masses $\hz$ and $\hz^*$,
 as well as a conjugate pair of bosonic quarks, with masses $\hz'$ and 
$\hz'^*$ \cite{Toublan:1999hx}. The graded mass term becomes
\be
\label{gradedM}
{\rm Trg}\big({\cal M}U+{\cal M}U^{-1}\big) \quad {\rm with} \quad
{\cal M}={\rm diag}(\hm_1,\ldots,\hm_{N_f},\hz,\hz^*,\hz',\hz'^*),
\ee
where Trg denotes the graded trace ${\rm Trg}A=\Tr(A_f)-\Tr(A_b)$, with $A_f$
the fermion-fermion block of $A$ and $A_b$ its boson-boson block.  
The eigenvalue density of $D_W$ in the complex plane is
\be
\rho_{c,N_f}^\nu(\hz,\hz^*,\hm;\ha_i) 
& = & \partial_{\hz^*}\lim_{\hz'\to\hz}\partial_{\hz} \log Z_{N_f+2|2}^\nu(\hz,\hz^*,\hz',\hz'^*,\hm;\ha_i).
\ee

The sign and magnitude of $W_6$, $W_7$ and $W_8$ determine the 
phase structure at small mass \cite{SharpeSingleton}: for $W_8+2W_6>0$ 
the Aoki phase dominates if $|m|\Sigma<8(W_8+2W_6)a^2$ while for $W_8+2W_6<0$ 
the Sharpe-Singleton scenario takes place. It is therefore of considerable 
interest to understand if it is possible to determine the signs of
the additional low energy constants. In the next section we show how these 
signs follow from the $\gamma_5$-Hermiticity of the Wilson Dirac operator.

\section{Constraints on $W_6$, $W_7$ and $W_8$ due to $\gamma_5$-Hermiticity}
\label{sec:constraints}

In Refs.~\cite{ADSVprd,HS1,SVtwist} it was shown that properties of the 
partition 
function and the correlation functions due to $\gamma_5$-Hermiticity lead 
to bounds on $W_6$, $W_7$ and $W_8$. The bounds that where found are 
\cite{ADSVprd,HS1} 
$W_8>0$ (independent of the value of $W_6$ and $W_7$ \cite{HS1}) 
and \cite{ADSVprd,SVtwist} $W_8-W_6-W_7>0$. In addition it was 
argued in \cite{SVtwist} that $W_8+2W_6>0$ provided that disconnected 
diagrams are suppressed. Note that lattice studies \cite{arXiv:0709.4564} 
have found that disconnected diagrams can have a significant contribution.

Here we show that the signs of $W_6$ and $W_7$ 
can be determined from $\gamma_5$-Hermiticity if we consider the spectral 
properties of the Wilson Dirac operator. There are two implicit assumptions
that have been well established in the study of Dirac spectra. 
First, that for a given value of the low-energy constants the chiral 
Lagrangian can be extended to partially quenched QCD with the same 
low-energy constants. Second, there is a one-to-one relation between 
spectral properties in the microscopic domain and the partially 
quenched chiral Lagrangian.

\vspace{3mm}

Let us first recall why $\gamma_5$-Hermiticity implies that $W_8>0$ when 
$W_6=W_7=0$ \cite{ADSVprd}. As shown by explicit calculations 
in \cite{DSV,ADSVprd,ADSVNf1,SV-Nf2} the microscopic graded 
generating functional corresponding to
\be\label{L-Wilson8}
{\cal L}(U) & = &
\frac 12 m \Sigma { \rm Tr} (U+U^\dagger)
+\frac 12 \zeta \Sigma{ \rm Tr} (U-U^\dagger) 
- a^2 W_8{\rm   Tr}(U^2+{U^\dagger}^2) 
\ee
with $W_8>0$ gives predictions for the spectrum of the 
$\gamma_5$-Hermitian $D_W$ and the Hermitian $D_5$. This was 
further confirmed by its equivalence to a $\gamma_5$-Hermitian 
Wilson Random Matrix Theory. 

On the contrary if $W_8<0$, it was explicitly shown 
in \cite{ADSVprd} that the graded generating functional 
corresponding to Eq. (\ref{L-Wilson8}) is the generating 
functional for the spectral fluctuations in a lattice 
theory with iWilson fermions defined as
\be
D_{iW} =\frac{1}{2}\gamma_\mu(\nabla_\mu+\nabla_\mu^*)
     -i\frac{ar}{2}\nabla_\mu\nabla_\mu^*,
\ee
which is anti-Hermitian rather than $\gamma_5$-Hermitian.
This conclusion was again confirmed by the equivalence to an 
anti-Hermitian iWilson Random Matrix Theory. 
Note that $D_W$ and $D_{iW}$ only differ by a factor of $i$ in 
the Wilson term, and that $D_{iW}$ is $not$ $\gamma_5$-Hermitian.

Therefore we understand the effective theory, 
Eq. (\ref{L-Wilson8}), for both signs of $W_8$ and 
that the Hermiticity properties of the Wilson Dirac operator 
determine this sign. For Wilson fermions we have $W_8>0$, 
whereas for iWilson fermions the constraint is $W_8< 0$.
This is fully consistent with the results from QCD inequalities 
\cite{ADSVprd,HS1}.

\begin{center}
\begin{figure}[t*]
\includegraphics[width=8cm,angle=0]{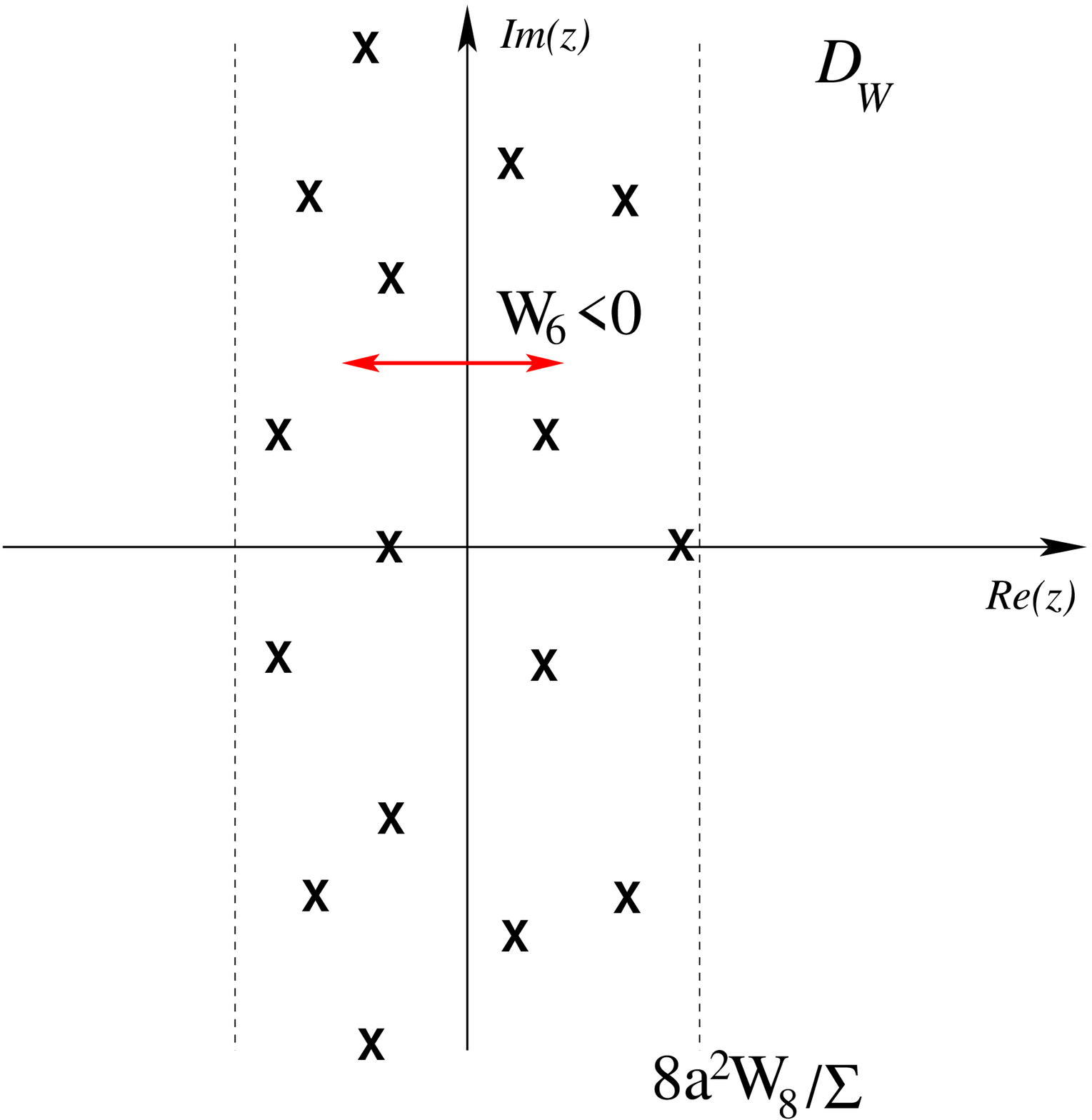}
\hfill
\includegraphics[width=8cm,angle=0]{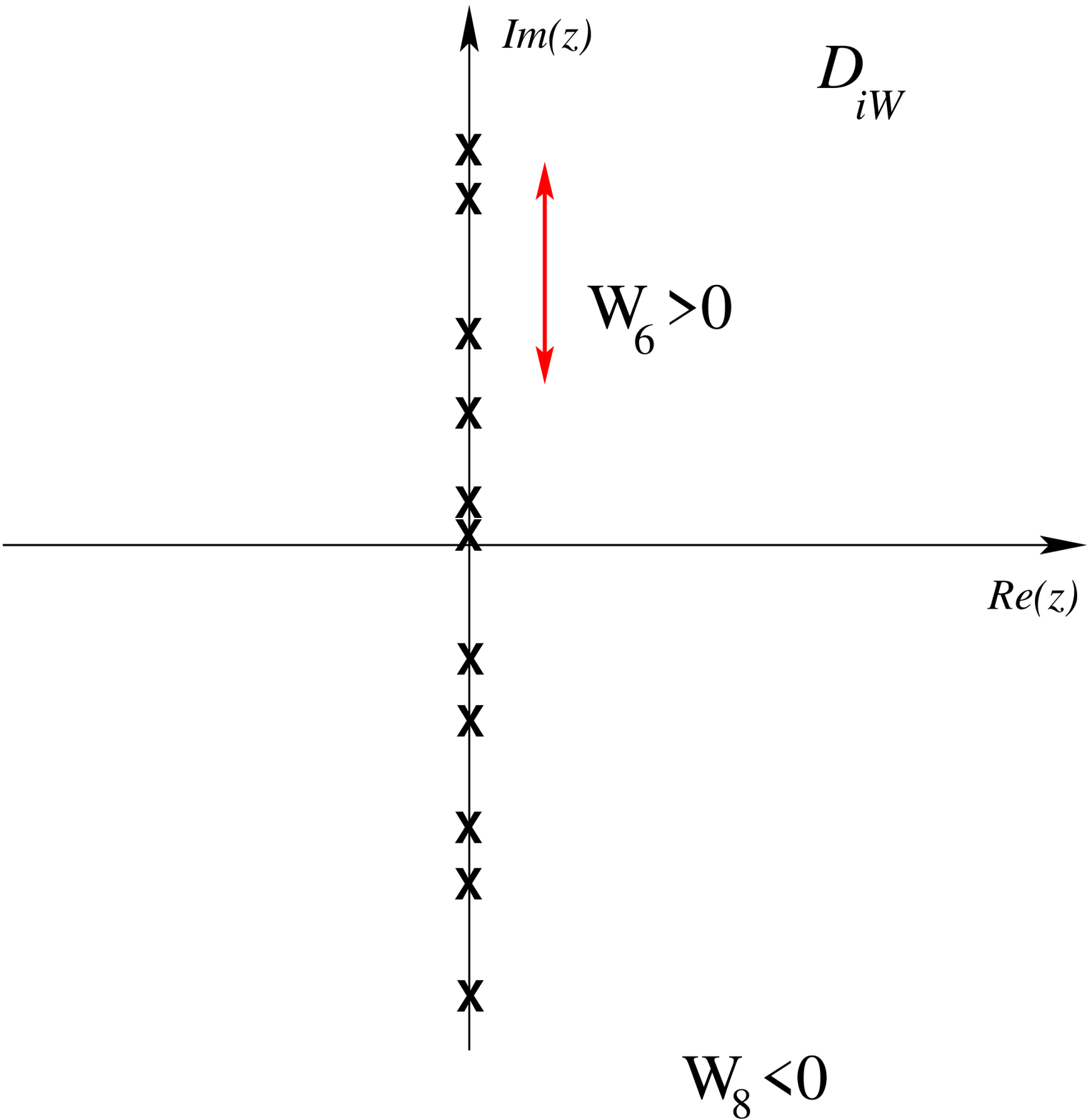}
\caption{\label{fig:DWandDiW} Illustration of the fluctuations 
of the Dirac eigenvalues. {\bf Left:} A negative value of $W_6$ 
corresponds to a $\gamma_5$-Hermitian Wilson 
Dirac operator, i.e. with eigenvalues that are either real 
or come in complex conjugate pairs. {\bf Right:} The Dirac operator 
corresponding to $W_6>0$ is in the Hermiticity class 
of $D_{iW}$ with purely imaginary eigenvalues.}
\end{figure}
\end{center}

Let us now extend the argument to also include $W_6$ and $W_7$.
We will show that Wilson chiral perturbation theory with 
$W_6<0$, $W_7<0$ and $W_8>0$  gives predictions for the spectrum 
of a $\gamma_5$-Hermitian $D_W$. 
On the contrary  Wilson chiral perturbation theory with 
$W_6>0$, $W_7>0$ and $W_8<0$ gives predictions for the spectrum 
of $D_{iW}$.

The fact that all three signs are reversed when changing between Wilson and 
iWilson fermions is not accidental.
Since the Wilson term and the iWilson term break 
chiral symmetry in exactly the same way, the respective low energy 
effective theories, must have the 
same symmetry breaking terms in the chiral Lagrangian. 
Moreover, since the explicit symmetry breaking terms at order $a^2$ have 
their origin in the Wilson term, the two effective fermionic 
Lagrangians 
are related by a combined 
change of sign of $W_6$, $W_7$ and $W_8$ \footnote{This duality of 
the Wilson and iWilson fermion lattice theories is due to the 
fact that the two are related by an axial transformation and an 
interchange $m\leftrightarrow i\zeta$ and 
$\zeta\leftrightarrow im$: The axial transformation $R=\exp(i\pi/4)$ 
and $L=\exp(-i\pi/4)$ takes $D_W+m+\zeta\gamma_5 
\to D_{iW}+i\zeta+im\gamma_5$. The corresponding transformation 
on the Goldstone field is $U \to R U L^\dagger = iU$.}.

In order to see which sign of $W_6$ and $W_7$ corresponds to Wilson 
fermions let us rewrite the trace squared terms in Wilson 
chiral perturbation theory as 
\be
Z^\nu_{N_f}(\hm,\hze;\ha_6,\ha_7,\ha_8) 
&=& \frac{1}{16 \pi |\ha_6 \ha_7|}\int_{-\infty}^\infty dy_6 dy_7 \ 
\exp\left[-\frac{y_6^2}{16|\ha_6^2|}-\frac{y_7^2}{16|\ha_7^2|}\right] \ 
\nn\\ && \times Z_{N_f}^\nu(\hm-y_6,\hze-y_7;\ha_6=0,\ha_7=0,\ha_8) ,   
\label{zrandmass}
\ee
valid for $W_6<0$ and $W_7<0$ and
 \be
Z^\nu_{N_f}(\hm,\hze;\ha_6,\ha_7,\ha_8) 
&=& \frac{1}{16 \pi |\ha_6 \ha_7|}\int_{-\infty}^\infty dy_6 dy_7 \ 
\exp\left[-\frac{y_6^2}{16|\ha_6^2|}-\frac{y_7^2}{16|\ha_7^2|}\right] \ 
\nn \\ &&\times
Z_{N_f}^\nu(\hm-iy_6,\hze-iy_7;\ha_6=0,\ha_7=0,\ha_8),   
\label{zrandimmass}
\ee 
valid for $W_6>0$ and $W_7>0$.

Let us first consider the case $W_7=0$. A negative value of $W_6$ 
corresponds to a Dirac operator that is compatible with the 
$\gamma_5$-Hermiticity of the Wilson 
Dirac operator. The additional fluctuations can be interpreted as
collective fluctuations of the eigenvalues, 
$z_k$, of $D_W$ parallel to the real $z$-axis. To see this, extend 
Eq.~(\ref{zrandmass}) to the graded generating functional, 
Eq.~(\ref{ZnuNf+2|2}), and include $y_6$ in the graded mass matrix
\be
{\cal M}-y_6={\rm diag}(\hm_1-y_6,\ldots,\hm_{N_f}-y_6,\hz-y_6,\hz^*-y_6,\hz'-y_6,\hz'^*-y_6)
\ee
(see Eq.~(\ref{rhoc_micro_a6a8}) below for further details). Such fluctuations 
are allowed for Wilson fermions since the eigenvalues of $D_W$ come 
in pairs $(z,z^*)$ or are strictly real. This is illustrated in the 
left hand panel of figure \ref{fig:DWandDiW}.

For a positive value of $W_6$ the corresponding Dirac operator
is in a different Hermiticity class than the Wilson Dirac operator 
and will have different spectral properties. Therefore, we necessarily 
have $W_6<0$ for the Wilson Dirac operator. 
For the iWilson-lattice theory on the other hand, we have that 
$D_{iW}^\dagger=-D_{iW}$ and consequently purely imaginary eigenvalues. 
Moreover, since the eigenvalues are {\sl not} paired with equal and 
opposite sign (for $a\neq0$) the spectrum of $iD_W$ can 
fluctuate along the imaginary axis, see the right hand panel 
of figure \ref{fig:DWandDiW} for an illustration. The Dirac operator 
corresponding to $W_6>0$ is hence in the Hermiticity class of $D_{iW}$. 
In perfect agreement with the 
above conclusion for Wilson fermions and the fact that the two effective 
theories should have opposite signs for all three $W_i$'s.

The story for $W_7$ is analogous: 
A negative value of $W_7$ corresponds to real fluctuations of the axial 
quark mass, which are compatible with the Hermiticity properties of the 
Wilson Dirac operator. These fluctuations can be interpreted as collective 
fluctuations of the eigenvalues, 
$\lambda^5$, of $D_5\equiv\gamma_5(D_W+m)$ parallel to the real 
$\lambda^5$-axis. 
Such fluctuations are allowed for Wilson fermions since $D_5$ is 
Hermitian and the symmetry $(\lambda^5,-\lambda^5$) is violated 
when $a\neq0$.

For iWilson fermions the product $\gamma_5D_{iW}$ has complex eigenvalues 
which come in pairs with opposite real part (or are strictly 
imaginary), hence their fluctuations can only take part in the 
imaginary direction. This is consistent with $W_7>0$ in the chiral 
Lagrangian for iWilson fermions and in perfect agreement with the fact 
that this sign should be opposite to that of the chiral Lagrangian 
for Wilson fermions.

Finally, when $W_6$ and $W_7$ have opposite signs the 
Hermiticity properties of the shifted Dirac operator always differ 
from the one realized at $W_6=W_7=0$. The corresponding Dirac operator 
therefore is neither $\gamma_5$-Hermitian nor anti-Hermitian. The same 
is true if all $W_i$ have the same sign.

\vspace{3mm}

In conclusion, we explained that
the signs of the low energy constants of Wilson chiral 
perturbation theory follow from the $\gamma_5$-Hermiticity of 
the Wilson Dirac operator. We have, $W_6<0$, $W_7<0$ and $W_8>0$. 
Note that both the Aoki phase with $W_8+2W_6>0$ and the Sharpe-Singleton 
scenario with $W_8+2W_6<0$ are allowed by $\gamma_5$-Hermiticity.

\vspace{2mm}

In the reminder of this paper we will work with $W_6<0$, $W_7<0$ and $W_8>0$. 
Moreover, since the low energy constant $W_7$ does 
not affect the competition between the Aoki phase and the Sharpe-Singleton 
scenario we will set $W_7=0$. 

\vspace{2mm}

In section \ref{sec:SharpeSingleton} below we show how a collective 
effect on the eigenvalues of $D_W$ induced by $W_6<0$ leads to a shift 
between the Aoki and the Sharpe-Singleton scenario. To establish this 
result we will first derive the unquenched microscopic eigenvalue 
density of $D_W$.

\section{The unquenched spectrum of $D_W$}
\label{sec:rho_c}

In this section we calculate the microscopic spectral density of the Wilson 
Dirac operator, $D_W$, in the presence of two dynamical flavors. We 
first carry through the calculation with $W_6=W_7=0$ and subsequently 
introduce the effects of $W_6$. In order to derive the microscopic 
spectral density of $D_W$ it is convenient to use Wilson chiral 
random matrix theory introduced in \cite{DSV}, which is reviewed
in \ref{app:WRMT} for completeness.

We start from the joint eigenvalue probability distribution of the 
random matrix 
partition function Eq.~(\ref{ZWRMTev}). To obtain the eigenvalue density 
in the complex plane we integrate over all but a complex pair of eigenvalues.
Using the 
properties of the Vandermonde determinant we obtain ($\hz=\hx+i\hy$)
\be
\label{rhocNf2}
\rho_{c,N_f=2}^\nu(\hz,\hz^*,\hm;\ha_8) & = & e^{-\hx^2/(8\ha_8^2)} 
 \frac{|\hy| e^{-4\ha_8^2}}{16(2\pi)^{5/2}2\ha_8} (\hz-\hm)^2(\hz^*-\hm)^2 
\frac{Z_4^\nu(\hz,\hz^*,\hm,\hm;\ha_8)}{Z_2^\nu(\hm,\hm;\ha_8)}. \nn \\
\ee
This amazingly compact form can be simplified further.
In \cite{K} it was shown that the four flavor partition function $Z_4^\nu$ 
can be expressed in terms of  two flavor partition functions. A 
 proof in terms of chiral Lagrangians is given in \ref{app:ZNf}. 
This leads to the final form for the microscopic spectral density 
of $D_W$ with two dynamical flavors 
\be
\label{rhocNf2fac}
\rho_{c,N_f=2}^\nu(\hz,\hz^*,\hm;\ha_8) & = & e^{-\hx^2/(8\ha_8^2)} 
 \frac{|\hy|e^{-4\ha_8^2}}{16(2\pi)^{5/2}2\ha_8}  Z_2^\nu(\hz,\hz^*;\ha_8) \\
&& \hspace{-1.3cm}\times \Big(1 - \frac{1}{2i\hy} \frac{\partial_{\hm}[\hat{Z}_2^\nu(\hz,\hm;\ha_8)]\hat{Z}^\nu_2(\hz^*,\hm;\ha_8)-\hat{Z}^\nu_2(\hz,\hm;\ha_8)\partial_{\hm}[\hat{Z}^\nu_2(\hz^*,\hm;\ha_8)]}{Z_2^\nu(\hm,\hm;\ha_8)Z^\nu_2(\hz,\hz^*;\ha_8)}\Big) \nn ,
\ee
where the two flavor partition function is given by \cite{SV-Nf2}
\be
Z^\nu_{N_f=2}(\hm_1, \hm_2;\ha_8) &= &\frac {e^{4\ha_8^2}}{\pi 8\ha_8^2}
\int_{-\infty}^\infty\int_{-\infty}^\infty ds_1 ds_2
\frac{(is_1-is_2)}{\hm_1-\hm_2} (is_1)^\nu(is_2)^\nu\tilde Z^\nu_2(is_1,is_2;\ha_8=0) \nn \\
&& \hspace{3cm} \times \exp\left[-\frac 1{16\ha_8^2}[(s_1+i\hm_1)^2+ (s_2+i\hm_2)^2 ]\right],
\label{ztwo}
\ee
with
\be
\tilde Z^\nu_2(x_1,x_2;\ha_8=0) = \frac 2{x_1^\nu x_2^\nu (x_2^2-x_1^2)} \det \left |
\begin{array}{cc} I_\nu(x_1) & x_1 I_{\nu+1}(x_1)
\\ I_\nu(x_2)& x_2 I_{\nu+1}(x_2)
\end{array}  \right |, 
\ee
and we have introduced the notation 
$\hat{Z}_2^\nu(\hm_1,\hm_2;\ha_8)\equiv(\hm_1-\hm_2)Z_2^\nu(\hm_1,\hm_2;\ha_8)$.

\begin{center}
\begin{figure}[t*]
\includegraphics[width=8cm,angle=0]{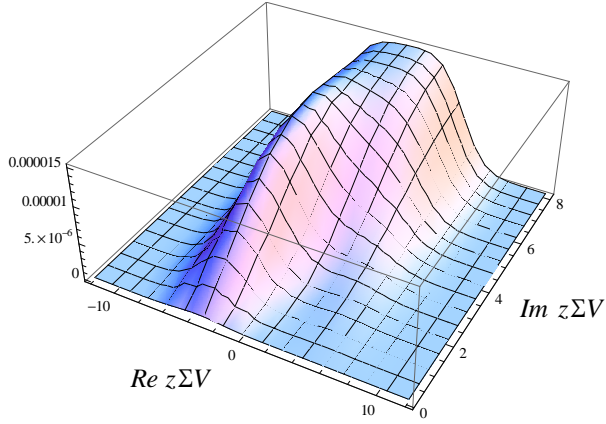}
\caption{\label{fig:rhocNf2} The microscopic spectral density of 
the Wilson Dirac operator for $N_f=2$ flavors of equal mass $\hm=2$ 
and $\ha_8=0.8$ ($\ha_6=\ha_7=0$) in the sector $\nu=0$. The eigenvalues 
form a strip centered on the imaginary axis. Note the 
repulsion of the eigenvalues from the quark mass.}
\end{figure}
\end{center}

The expression in the first line of Eq. (\ref{rhocNf2fac}) is the 
quenched eigenvalue density of $D_W$ \cite{KVZ}. The correction 
factor in the second line is 
responsible for the eigenvalue repulsion from the quark mass. 
A plot of the eigenvalue density of the Wilson Dirac 
operator in the complex plane for two dynamical flavors
is given in figure \ref{fig:rhocNf2}. 

Note the strong similarity with the result for the eigenvalue density 
of the continuum Dirac operator at nonzero chemical potential in phase 
quenched QCD \cite{AOSV}. In that case the eigenvalue density follows 
from the integrable Toda lattice hierarchy \cite{Toda}. The analytical  
form of the eigenvalue density of the Wilson Dirac operator, Eq. 
(\ref{rhocNf2}), strongly suggests that a similar integrable structure 
is present in the microscopic limit of the Wilson lattice QCD partition 
function.

\subsection{Including the effect of $W_6$}

As pointed out in \cite{ADSVprd} the graded 
generating function for the eigenvalue density can be extended to 
include the effect of $W_6$ and $W_7$ by a Gaussian integral as in 
Eq.~(\ref{zrandmass}). 
Since this works for the graded generating functional it also works 
for the spectral density itself \cite{ADSVprd}. 
In the unquenched case, however, 
one must be careful with the normalization factor $1/Z_{N_f}^\nu(\hm;\ha_8)$. 

Let us start with the case where $W_6=W_7=0$. Then the density of $D_W$ in 
the complex plane is obtained from the graded generating 
function as follows 
\be
\rho_{c,N_f}^\nu(\hz,\hz^*,\hm;\ha_8) 
& = &{ \partial_{\hz^*}\Sigma_{N_f+2|2}^\nu(\hz,\hz^*,\hm;\ha_8)
} \nn\\
 & = & \partial_{\hz^*}\lim_{\hz'\to\hz}\partial_{\hz} \log Z_{N_f+2|2}^\nu(\hz,\hz^*,\hz',\hz'^*,\hm;\ha_8) \ ,
\ee
where the graded generating functional, $Z_{N_f+2|2}$, was introduced 
in Eq.~(\ref{ZnuNf+2|2}). 

To extend this to $W_6<0$ we first note that the Gaussian trick, 
Eq.~(\ref{zrandmass}), also works for the graded generating functional. 
Using this we find
\be
\label{rhoc_micro_a6a8}
\rho_{c,N_f}^\nu(\hz,\hz^*,\hm;\ha_6,\ha_8) 
& = & \partial_{\hz^*}\lim_{\hz'\to\hz}\partial_{\hz} \log Z_{N_f+2|2}^\nu(\hz,\hz^*,\hz',\hz'^*,\hm;\ha_6,\ha_8) \\
& = & \partial_{\hz^*}\lim_{\hz'\to\hz}\partial_{\hz} \log\int [dy] Z_{N_f+2|2}^\nu(\hz-y,\hz^*-y,\hz'-y,\hz'^*-y,\hm-y;\ha_8) \nn \\
& = & \frac{1}{Z_{N_f}^\nu(\hm;\ha_6,\ha_8)}\int [dy]  \ Z_{N_f}^\nu(\hm-y;\ha_8) \ \partial_{\hz^*} \Sigma_{N_f+2|2}^\nu(\hz-y,\hz^*-y,\hm-y;\ha_8) \nn\\
& = & \frac{1}{Z_{N_f}^\nu(\hm;\ha_6,\ha_8)}\int [dy] \ Z_{N_f}^\nu(\hm-y;\ha_8)\rho_{c,N_f}^\nu(\hz-y,\hz^*-y,\hm-y;\ha_8) \nn,
\ee
where we will recall the notation: $[dy]=dy/(4\sqrt{\pi}|\ha_6|)\exp(-y^2/(16|\ha_6^2|))$.

In order to understand the effect of $W_6$ on the unquenched spectral 
density of $D_W$ we will analyze the mean field limit of 
Eq.~(\ref{rhoc_micro_a6a8}). 
As is shown in the next section the factor of $Z_{N_f}^\nu$ in the 
integrand, is essential for the realization of the Sharpe-Singleton scenario.

\section{The Sharpe-Singleton Scenario in the spectrum of $D_W$}
\label{sec:SharpeSingleton}

Here we show that the Sharpe-Singleton scenario can  be understood in terms of
a collective 
effect of the eigenvalues of $D_W$ induced by $W_6<0$  when 
the quark mass changes sign. The Sharpe-Singleton scenario 
is therefore not realized in the quenched theory even if $W_8+2W_6<0$.

\vspace{3mm}

Before we give the proof let us first consider an electrostatic 
analogy which can help set the stage. The quenched chiral condensate 
\be
\int d^2 z  \ \frac{\rho_{N_f=0}(z,z^*;a)}{z-m}
\ee
can be thought of as the electric field (in two dimensions) created by 
positive charges located at the positions of the eigenvalues $z$ of 
$D_W$ and measured at the position $m$ (which can be thought of as a 
test charge). At the point where 
the quark mass hits the strip of eigenvalues of 
$D_W$ centered on the imaginary axis, the mass dependence
of the chiral condensate (electric 
field) shows a  kink. 
As the quark mass is lowered further 
(the test charge passes through the strip of eigenvalues) the condensate 
(electric field) drops linearly to zero at $m=0$. The drop is linear 
because the eigenvalue density is uniform.

For the unquenched chiral condensate we reach an identical conclusion provided 
that the quark mass (test charge) only has a local effect on the 
eigenvalues, i.e. it only affects eigenvalues close to the quark 
mass. This is the case for the Aoki phase when the quark mass is inside
the strip of 
 eigenvalues of $D_W$.

On the contrary, in order to realize the first order Sharpe-Singleton 
scenario the quark mass must have a collective effect on the eigenvalues 
of $D_W$ such that the strip of eigenvalues is entirely to the left of 
the quark mass for small positive values of $m$ and then at $m=0$ 
the strip collectively jumps to the opposite side of the origin such 
that for small negative values of the quark mass the strip of 
eigenvalues is to the right of $m$. The collective jump of the 
eigenvalues at $m=0$ flips the sign of the chiral condensate 
(electric field) in agreement with the Sharpe-Singleton 
scenario.

In order to show that the Sharpe-Singleton scenario is indeed 
realized in terms of the eigenvalues of $D_W$ in the manner described above 
let us analyze the effect of $W_6<0$ on the eigenvalues of $D_W$.

\subsection{The mean field eigenvalue density of $D_W$}

In the mean field limit the density of eigenvalues of $D_W$ at 
$\ha_6=0$ is simply given by a uniform strip of half width 
$8\ha_8^2/\Sigma$ centered on the imaginary axis (the deriviation 
of this result is analogous to the one for nonzero chemical potential, 
see \cite{Toublan:1999hx,Osborn:2008ab})
\be
\rho_{c,N_f=2}^{\rm MF}(\hx,\hm;\ha_8)=\theta(8\ha_8^2-|\hx|).
\ee
This result is identical to the quenched mean field spectral density 
since the correction factor in the second line of Eq. 
(\ref{rhocNf2fac}) only has an effect on the microscopic scale (the direct 
repulsion of the eigenvalues from the quark mass has a microscopic range).  

To include the effect of $\ha_6$ we use the Gaussian trick  discussed in
Eq. (\ref{rhoc_micro_a6a8}). The simplest way to proceed is to take the 
mean field limit before the $y_6$-integration, we find
\be
\rho_{c,N_f=2}^{\rm MF}(\hx,\hm;\ha_6,\ha_8)=\frac{1}{Z_2^{\rm MF}(\hm;\ha_6,\ha_8)} 
\int dy_6 \ e^{-y_6^2/16|\ha_6^2|}Z_2^{\rm MF}(\hm-y_6;\ha_8)\theta(8\ha_8^2-|\hx-y_6|).
\nn \\
\ee
Note the essential way in which the two flavor partition function 
enters both in numerator and the denominator. This is what separates 
the mean field calculation with dynamical fermions from the quenched analogue.

The mean field result for the two flavor partition function with 
$\ha_6=0$ is given by
\be
\label{ZNf2MFa8}
Z_2^{\rm MF}(\hm;\ha_8) = e^{2\hm-4\ha_8^2}+e^{-2\hm-4\ha_8^2}+\theta(8\ha_8^2-|\hm|)e^{\hm^2/8\ha_8^2+4\ha_8^2}.
\ee
The $\ha_6$ dependence can again be restored  by means of introducing
an additional Gaussian integral. 
In the mean field limit this results in 
\be
Z_2^{\rm MF}(\hm;\ha_6,\ha_8) & = & 
e^{2\hm+16|\ha_6^2|-4\ha_8^2}+e^{-2\hm+16|\ha_6^2|-4\ha_8^2} \\
&&+\theta(8(\ha_8^2+2\ha_6^2)-|\hm|)e^{\hm^2/8(\ha_8^2-2|\ha_6^2|)+4\ha_8^2}.
\nn
\ee
Note that when $2\ha_6^2 +\ha_8^2 <0$ the term in the second line of this 
equation is absent. The final result for the mean field two flavor eigenvalue 
density of $D_W$ is
\be
\label{rhocNf2MF}
\rho_{c,N_f=2}^{\rm MF}(\hx,\hm;\ha_6,\ha_8) & = & \frac{1}{Z_2^{\rm MF}(\hm;\ha_6,\ha_8)} 
\\ &&
\hspace{-1.3cm}\times \Big\{ e^{2\hm+16|\ha_6^2|-4\ha_8^2}\theta(8\ha_8^2-|\hx+16|\ha_6|^2|)\nn\\ &&
\hspace{-1cm}+e^{-2\hm+16|\ha_6^2|-4\ha_8^2}\theta(8\ha_8^2-|\hx-16|\ha_6|^2|)
\nn\\
&&\hspace{-1cm}+\theta(8(\ha_8^2+2\ha_6^2)-|\hm|)
\theta\left(8\ha_8^2-\left|\hx+\frac{2|\ha_6|^2\hm}{(\ha_8^2-2|\ha_6|^2)}\right|\right)
e^{\hm^2/8(\ha_8^2-2|\ha_6^2|)+4\ha_8^2}\Big\}\nn.
\ee
A derivation of this result which includes the fluctuations around the 
saddle points is given in \ref{app:MF}.

In order to access the Sharpe-Singleton scenario let us consider 
the case where $\hm$ is small compared to $16|\ha_6^2|-8\ha_8^2$ which is 
taken large and positive.  

\begin{center}
\begin{figure}[t*]
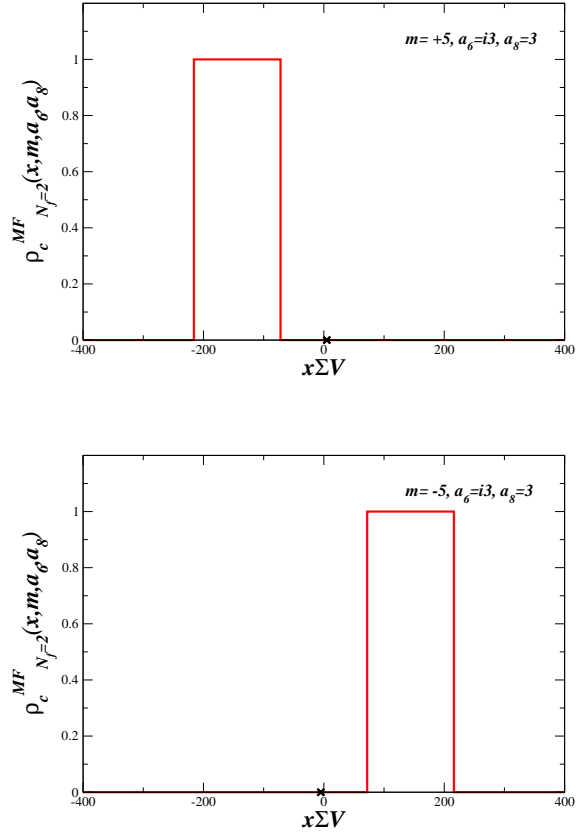

\includegraphics[width=7.5cm,angle=0]{m5.eps}
\\\vspace{1cm}
\includegraphics[width=7.5cm,angle=0]{mm5.eps}
\caption{\label{fig:rhocNf2MF} The Wilson Dirac spectrum
for the Sharpe-Singleton 
scenario: Shown is the mean field spectral density of the Wilson Dirac 
operator for $N_f=2$ with $\ha_6=3i$ and $\ha_8=3$ ($\ha_7=0$) as a 
function of $\hx={\rm Re}[\hz]$ (the mean field density is independent of 
$\hy={\rm Im}[\hz]$). The choice of $\ha_6$ and $\ha_8$ corresponds to 
a negative value of $W_8+2W_6$ and hence the Sharpe-Singleton scenario. 
The two flavors have equal mass $\hm=5$  ({\bf top}) and $\hm=-5$ ({\bf bottom}). Even though the quark mass, marked by {\bf x} on the $x$-axis, 
only changes by a small amount compared to the size of the gap 
the entire strip of eigenvalues jumps to the opposite side of the 
origin. This leads to the first order jump of the chiral condensate at 
$m=0$.}
\end{figure}
\end{center}

The terms in the second line of Eq. (\ref{rhocNf2MF}) give rise to a strip 
of eigenvalues of half width 8$\ha_8^2/\Sigma$ centered at -16$|\ha_6^2|/\Sigma$
while the term in the third line gives rise to a strip 
of eigenvalues of half width 8$\ha_8^2/\Sigma$ centered at 16$|\ha_6^2|/\Sigma$.
The relative height of the two strips is $\exp(4\hm)$. Therefore
even though the magnitude of $\hm$ is relatively small it has a 
dramatic effect: As the sign of $\hm$ changes from positive to 
negative values the entire strip of eigenvalues jumps from 
its position around -16$|\ha_6^2|/\Sigma$ to the new position around 
16$|\ha_6^2|/\Sigma$. For a plot see figure \ref{fig:rhocNf2MF}.
Because of the exponential suppression of one of the strips, the jump
of the support of the spectrum
occurs on a scale of $\hm \sim O(1)$ or $m \sim 1/V\Sigma$ and 
leads to the first order discontinuity of the chiral 
condensate at $m=0$ as predicted by the Sharpe-Singleton scenario.

In the continuum limit the chiral condensate also jumps from $\Sigma$ to
$ -\Sigma$ on a scale of   $\hm \sim O(1)$ or $m\sim 1/V\Sigma$, but in this
case the difference in the potential between the two minima is of $O(\hm)$ 
as opposed to $O(\ha_6^2)$ for the Sharpe-Singleton scenario.

The terms in the mean field two flavor partition function, 
see Eq.~(\ref{ZNf2MFa8}), 
are directly responsible for the jump of the eigenvalue density at 
$\hm=0$ in the theory with dynamical quarks. 
In the corresponding quenched computation
we simply have 
\be
\rho_{c,N_f=0}^{\rm MF}(\hx;\ha_6,\ha_8)=\int dy_6 \ e^{-y_6^2/16|\ha_6^2|}\theta(8\ha_8^2-|\hx-y_6|),
\ee
which leads to a single strip of eigenvalues centered at the 
imaginary axis independent of the value of $W_6$ 
\be
\rho_{c,N_f=0}^{\rm MF}(\hx;\ha_6,\ha_8)=\theta(8\ha_8^2-|\hx|).
\ee

\subsection{The connection to the mean field results of 
Sharpe and Singleton}

From the results of the previous subsection we see that the 
gap from the quark mass to the edge of the strip of eigenvalues 
of $D_W$ is given by
\be
|m|-8(W_8+2W_6)a^2/\Sigma.
\ee
In \cite{SharpeSingleton} it was found that the pion masses for
   $|m|\Sigma>8(W_8+2W_6)a^2$ are given by
\be
\frac{m_\pi^2F_\pi^2}{2} = |m|\Sigma-8(W_8+2W_6)a^2.
\ee
Hence the gap from the quark mass to the edge of the strip of
eigenvalues of $D_W$ can be thought of as the effective quark
mass that enters the standard form of the GOR-relation. 
In particular, note that for $W_8+2W_6<0$ the mass never 
reaches the strip of eigenvalues. Correspondingly, the minimal 
value of the pion mass is given by 
\be
\frac{m_\pi^2F_\pi^2}{2} = -8(W_8+2W_6)a^2,
\ee
again in perfect agreement with the leading order 
$p$-regime computation of \cite{SharpeSingleton}.

\subsection{Direct computation of the quenched and unquenched condensate}

\begin{center}
\begin{figure}[t*]
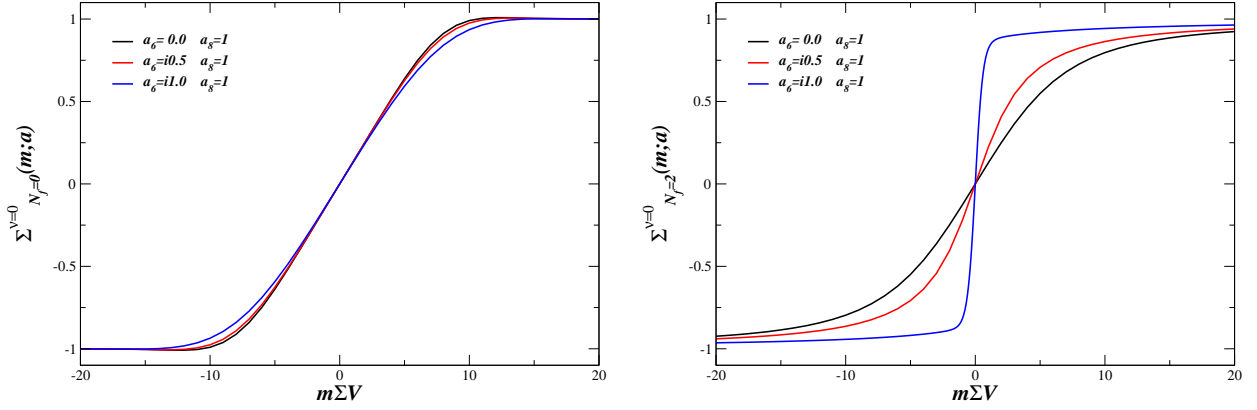

\includegraphics[width=8cm,angle=0]{Wilson-quenched-cond-nu0.eps}
\hfill
\includegraphics[width=8cm,angle=0]{Wilson-Nf2-cond-nu0.eps}
\caption{\label{fig:cond} The Sharpe-Singleton first order phase 
transition is due to dynamical quarks and is not present in the quenched case 
even if $W_8+2W_6<0$. Shown is the microscopic chiral condensate 
as a function of the quark mass for $\ha_8=1$ and $\ha_6=0,0.5i$ 
and $i$ corresponding to $W_8+2W_6>0$, $W_8+2W_6=0$ and $W_8+2W_6<0$,  
respectively. {\bf Left} $N_f=0$: In the quenched case there is hardly 
any effect of $W_6<0$. {\bf Right} $N_f=2$: For two flavors the 
increasingly negative $W_6$ drives the system from the Aoki phase 
to the Sharpe-Singleton scenario as can be seen by the formation of the 
discontinuity of the chiral condensate on a scale of $m \sim 1/V$.}
\end{figure}
\end{center}

From the essential part played by the dynamical fermion determinant in the
realization of the Sharpe-Singleton scenario in terms of the eigenvalues of 
the Wilson Dirac operator we conclude that the Sharpe-Singleton 
first order scenario only takes place in the theory with dynamical quarks. 
Here we explicitly compute the quenched and unquenched microscopic 
chiral condensate and directly verify that the first order jump of 
the chiral condensate at $m=0$ only takes place in the 
theory with dynamical quarks.

\hspace{3mm}

The unquenched microscopic chiral condensate is obtained from 
the microscopic partition function by
\be
\Sigma_{N_f}^\nu(\hm;\ha_i) = \frac{1}{N_f}\frac{1}{Z_{N_f}^\nu}\frac{d}{d\hm} 
Z_{N_f}^\nu(\hm;\ha_i).
\ee 
Specifically, for two mass degenerate flavors we have \cite{DSV}
\be
\Sigma_{N_f=2}^\nu(\hm,\hm;\ha_i) & = & \frac{1}{2} \frac{1}{Z_{N_f=2}^\nu(\hm;\ha_i)} 
\\ && \times \int_{-\pi}^\pi d\theta_1 d\theta_2 |e^{i\theta_1} - e^{i\theta_2}|^2  e^{i\nu(\theta_1 + \theta_2)} 
(\cos\theta_1 + \cos\theta_2) \nn\\
&& \hspace{-1.5cm} \times\exp\left[\hm(\cos\theta_1 + \cos\theta_2) - 4 \ha_6^2 (\cos\theta_1 + \cos\theta_2)^2 - 2\ha_8^2 (\cos(2\theta_1)+ \cos(2\theta_2))\right] \nn
\ee
with
\be
Z_{N_f=2}^\nu(\hm;\ha_i)& = & \int_{-\pi}^\pi d\theta_1 d\theta_2 |e^{i\theta_1} - e^{i\theta_2}|^2  e^{i\nu(\theta_1 + \theta_2)} \\
&& \hspace{-1.5cm} \times\exp\left[\hm(\cos\theta_1 + \cos\theta_2) - 4 \ha_6^2 (\cos\theta_1 + \cos\theta_2)^2 - 2\ha_8^2 (\cos(2\theta_1)+ \cos(2\theta_2))\right] . \nn
\ee

The quenched condensate was derived in \cite{DHS,ADSVprd} 
\be\label{SigmaQ2}
\Sigma^\nu_{N_f=0}(\hm;\ha_i) 
& = & \int_{-\infty}^\infty ds\int_{-\pi}^\pi \frac{d\theta}{2\pi} \ 
\sin(\theta)e^{(i\theta-s)\nu} 
\exp[-\hm\sin(\theta)- i \hm\sinh(s)-\epsilon \cosh s \nn \\
&&
\hspace{-1.0cm}+4\ha_6^2(-i\sin(\theta)+\sinh(s))^2
+4\ha_7^2(\cos(\theta)-\cosh(s))^2
+2\ha_8^2(\cos(2\theta)-\cosh(2s))] \nn \\
&& \hspace{-1.0cm}\times 
\Big(-\frac{\hm}{2}\sin(\theta)+i\frac{{\hm}}{2}\sinh(s)
-4(\ha_6^2+\ha_7^2)(\sin^2(\theta)+\sinh^2(s))
\nn\\
&& 
+2{\ha}_8^2(\cos(2\theta)+\cosh(2s)+e^{i\theta+s}+e^{-i\theta-s})+\frac{1}{2}\Big).
\ee
In Refs.~\cite{DHS,ADSVprd}, its imaginary part was studied since it is 
directly 
related to the real eigenvalues of $D_W$. Here we are after the quenched 
condensate itself which is given by its real part. 
Figure \ref{fig:cond} compares the behavior of the quenched chiral 
condensate and the chiral condensate for $N_f=2$ for three sets 
of values of $W_6$ and $W_8$. While the first order jump forms in the 
thermodynamic limit for the condensate with dynamical quarks when 
$W_8+2W_6$ turns negative the kink 
in the mass dependence of the quenched condensate remains. This directly 
verifies that the Sharpe-Singleton scenario is absent in
quenched theory independent of the value of $W_6$. 
  
Note that the authors of \cite{GSS} concluded that both the Aoki phase 
and the Sharpe-Singleton scenario are possible in the quenched theory. 
They reached this conclusion because they worked in the large $N_c$ 
limit in which $W_6$ and $W_7$ vanish, and because the constraint on 
the sign of $W_8$ was not known at the time. 

\section{Conclusions}
\label{sec:conc}

The first order scenario of Sharpe and Singleton for 
lattice QCD with Wilson fermions has been studied from the perspective
of the eigenvalues of the Wilson Dirac operator. 
The behavior of the Wilson Dirac eigenvalues not only gives
constraints on the additional low energy parameters of Wilson chiral
perturbation theory ($W_6<0$, $W_7<0$ and $W_8>0$), it also allows 
us to explain the way in which the first order discontinuity 
of the chiral condensate is realized. In particular, we have shown that 
the associated collective jump of the spectrum of the  Wilson Dirac operator only 
occurs in the theory with dynamical quarks. The Sharpe-Singleton scenario is 
therefore not realized in the quenched theory which enters in 
the Aoki phase at sufficiently small quark mass. 
By a direct computation of the 
quenched microscopic chiral condensate we verified
that the  second order phase transition occurs in the quenched theory 
even if $W_8+2W_6<0$. This explains the puzzle 
why the Aoki phase dominates in the chiral limit of quenched lattice 
simulations while both the Aoki phase 
and the Sharpe-Singleton scenario have been observed in 
lattice QCD with dynamical Wilson fermions.

The above conclusion was made possible by the computation of the 
exact analytical result for 
 the microscopic spectral density of the Wilson Dirac operator in 
lattice QCD with two dynamical flavors. The explicit form of the 
microscopic expression allowed us to compute the mean field eigenvalue 
density and in turn make a direct connection to the original 
leading order $p$-regime results of Sharpe and Singleton.

It would be most interesting to test the predictions presented in this paper 
against dynamical lattice QCD simulations. Since the effects of $W_6$ 
and $W_8$ on the spectrum of $D_W$ in the unquenched theory are 
drastically different this offers a direct way to determine the 
values of these low energy constants. An early lattice study 
of the Wilson Dirac eigenvalues in dynamical simulations with 
light quarks appeared in \cite{Farchioni:2002vn}.

Finally, since the additional low energy constants of Wilson chiral 
perturbation theory parameterize the discretization errors, it is also 
most interesting to consider the effects of improvements of the lattice 
action on the unquenched spectrum of the Wilson Dirac operator 
\cite{Hasenfratz:2007rf}.

\noindent
{\bf Acknowledgments:}
We would like to thank the participants of the ECT$^*$ workshop 
'Chiral dynamics with Wilson fermions' for useful discussions. 
This work was supported by the Alexander-von-Humboldt 
Foundation (MK), U.S. DOE Grant No. DE-FG-88ER40388 (JV) 
and the {\sl Sapere Aude} program of The Danish Council for 
Independent Research (KS).


\renewcommand{\thesection}{Appendix \Alph{section}}
\setcounter{section}{0}

\section{Wilson Random Matrix Theory}
\label{app:WRMT}

In order to derive the microscopic spectral density of $D_W$ it is 
convenient to use Wilson chiral random matrix theory introduced in 
\cite{DSV}.

The partition function of Wilson chiral random matrix theory 
is defined as
\be
\label{ZWRMT}
\tilde{Z}^\nu_{N_f} = \int dAdBdW \ \prod_{f=1}^{N_f}
\det(\tilde{D}_W+\tm_f) \ {\cal P}(A,B,W). 
\ee
The matrix integrals are over the complex Haar measure. 

The random matrix analogue of the Wilson Dirac operator is 
\be
\tilde{D}_W= \mat \tilde a A & iW \\ iW^\dagger & \tilde a B \emat,
\label{w-diracOP}
\ee
where  
\be
A=A^\dagger \quad {\rm and}  \quad B^\dagger = B
\ee
are $(n+\nu) \times (n+\nu)$ and $n \times n$ complex matrices, 
respectively, and $W$ is an arbitrary complex $(n+\nu)\times n$ matrix.
Finally, the weight is
\be
\label{P}
{\cal P}(A,B,W) \equiv \exp\left[-\frac {N}{4}{\rm Tr}[A^2+B^2] 
-\frac N2 {\rm Tr} [ W W^\dagger]\right],
\ee
where $N=2n+\nu$.

As was shown in Ref.~\cite{ADSVprd}, the Wilson random matrix partition function
matches the microscopic partition function of Wilson chiral perturbation 
theory in the limit $N\to\infty$ with $N\tilde{m}$ and $N\tilde{a}^2$ fixed provided that we identify 
\be
N\tm = {m\Sigma V},
\quad \quad \frac{N\ta^2}{4} = a^2 W_8 V.
\label{match}
\ee

An eigenvalue representation of the partition function was derived in 
\cite{KVZ} 
\be
\label{ZWRMTev}
\tilde{Z}^\nu_{N_f} & = & \int dZ \ \Delta_{2n+\nu}(Z) \ \prod_{a=1}^n(z_{ar}-m)^{N_f}\prod_{b=1}^{n+\nu}(z_{bl}-m)^{N_f} \prod_{a=1}^n g_2(z_{al},z_{ar})  \prod_{b=1}^\nu z_{bl}^{b-1}g_1(z_{bl})
\ee
where $Z=(z_{1r}\ldots,z_{nr},z_{1l},\ldots,z_{n+\nu,l})$ are the $2n+\nu$ 
eigenvalues of $D_W$ and 
\be
g_1(z)=\sqrt{\frac{n}{2\pi \tilde{a}^2}}\exp\left[-\frac{n}{2\tilde{a}^2}x^2\right]\delta(y),
\ee
and
\be
g_2(z_1,z_2) & = &
\sqrt{\frac{n^3}{4\pi \tilde{a}^2(1+a^2)}}\frac{z_1^*-z_2^*}{|z_1-z_2|}\nonumber\\
 &\times&\left[\exp\left[-\frac{n(x_1+x_2)^2}{4\tilde{a}^2}-\frac{n(y_1-y_2)^2}{4}\right]
\delta^{(2)}(z_1-z_2^*)
\right.\nonumber\\
 &+&\left.\frac{1}{2}\exp\left[-\frac{n}{4\tilde{a}^2}(x_1+x_2)^2+\frac{n}{4}(x_1-x_2)^2\right]\right.\nonumber\\
 &\times&\left.{\rm erfc}\left[\frac{\sqrt{n(1+\tilde{a}^2)}}{2\tilde{a}}|x_1-x_2|\right]\delta(y_1)\delta(y_2)\right].
\ee
Finally, $\Delta(Z)$ is the Vandermonde determinant of the 
$2n+\nu$ eigenvalues.

In section \ref{sec:rho_c} we use this eigenvalue representation to derive the 
general form of the unquenched spectral density of $D_W$.

\section{Simplification of the partition function}
\label{app:ZNf}

In this appendix we express the general partition function with 
even $N_f$ in terms of a Pfaffian of two flavor partition functions.
This Pfaffian form was first given in \cite{K}. Here we give a  proof
in terms of chiral Lagrangians rather than random matrix theories.
In particular, we explicitly express the four 
flavor partition function entering Eq. (\ref{rhocNf2}) in terms 
of  two flavor partition functions. 

We start from the general $N_f$ microscopic partition function, 
Eq. (\ref{Znu}), with $\ha_6=\ha_7=0$ and make use of the identity 
\be
\exp\left[\ha_8^2 {\rm Tr}(U^2+U^{-2})\right]&=&
\exp\left[2N_f \ha_8^2+ \ha_8^2 {\rm Tr}(U-U^{-1})^2\right],\nn \\
&=&c e^{2N_f\ha_8^2}\int d\sigma \exp\left[ \frac{{\rm Tr} \sigma^2}{16 \ha_8^2}+ \frac i2 {\rm Tr}\sigma (U-{U^{-1}})\right],
\label{hsreal}
\ee
where $\sigma$ is an $N_f\times N_f$ anti-Hermitian matrix and $c$ a 
normalization constant. After a  shift of $\sigma$ by ${\cal M}$ we obtain 
 \be
\label{Zreal1}
Z^\nu_{N_f}({\cal M};\ha_8)  & = & c e^{2N_f \ha_8^2}\int d\sigma \int \hspace{-1.5mm} dU \
{\rm det}^\nu(i U)  \exp\left[ \frac{{\rm Tr} (\sigma-{\cal M})^2}{16 \ha_8^2}
 + \frac i2 {\rm Tr}\,\sigma(U -U^{-1})\right].
\ee
The next step is to decompose $\sigma = u S u^{-1}$ with $S$ a diagonal matrix
and perform the integration over $u$ by the Itzykson-Zuber integral. 
We find
\be
Z^\nu_{N_f}({\cal M};\ha_8)
& =&  \frac{e^{2N_f \ha_8^2}}{(16\pi \ha_8^2)^{N_f/2}}\int ds \frac {\Delta(S)}{\Delta({\cal M})} \exp\left[\frac{{\rm Tr}(S-{\cal M})^2}{16\ha_8^2}\right]
 \nn\\  && 
  \times  \prod_{k} (is_k)^{\nu}
 \tilde Z_{N_f}^\nu\left ( \big\{is_k\big\}; \ha_8=0 \right).\hspace*{0.5cm}
\label{S_N_f}
\ee
The Vandermonde determinant is defined by
\be
\Delta(x_1,\cdots, x_p) = \prod_{k>l}^p(x_k-x_l),
\ee
and an explicit expression for the partition function at $\ha_8=0$ is given by
 \be
 && \hspace{-2cm}  \tilde Z_{N_f}^\nu(x_1, \cdots, x_{N_f};\ha_8=0) \\
 & = &  c 
\left (\frac{1} {\prod_{k=1}^{N_f} x_k}\right )^\nu
\frac{\det [ (x_k)^{l-1}I_{\nu+l-1}(x_k)]}
{\Delta(x_1^2,\cdots,x_{N_f}^2) }.
\nn 
\label{za=0}
\ee
We have that
\be
\Delta(x_k)\prod_k (x_k)^\nu  \tilde{Z}^\nu_{N_f}(x_k;\ha_8=0) =
\frac {\Delta(x_k)}{\Delta(x_k^2)}\det x_k^{l-1} I_{\nu+l-1}(x_k),
\ee
which we will denote by the symbol $D$. We now express $D$ as a Pfaffian. 

By using recursion relations for Bessel functions, $D$ can be rewritten as
\be
D\equiv\frac {\Delta(x_k)}{\Delta(x_k^2)}\det x_k^{l-1} I_{\nu+P(l-1)}(x_k),
\ee
where $P(k) = (1-(-1)^k)/2$. Writing the determinant as a sum over
permutations and splitting the permutations into permutations of
odd integers, $\pi^o$, even integers, $\pi^e$, and the mixed 
permutations of even and odd integers, $\pi^{eo}$, we obtain
\be
D\equiv\frac {\Delta(x_k)}{\Delta(x_k^2)}
\sum_{\pi^{eo}}(-1)^{\sigma^{eo}} \sum_{\pi^e}\sum_{\pi^o}(-1)^{\sigma^{e}+\sigma^o}
\prod_{ l=0\; {\rm odd}}^{n-1} x_{\pi^o(l)}^{2l} I_\nu(x_{\pi^o(l)})
\prod_{ l=0\; {\rm even}}^{n-1} x_{\pi^e(l)}^{2l+1}I_{\nu+1}(x_{\pi^e(l)}).
\ee
The permutation over the even and odd integers can be resummed into
a Vandermonde determinant
\be
\sum_{\pi^o}\prod_{ l=0\; {\rm odd}}^{n-1}(-1)^{\sigma^o} x_{\pi^o(l)}^{2l} I_\nu(x_{\pi^o(l)})
&=&\Delta(x^2_{k^o}) \prod_{k^o \; odd} I_\nu(x_{k^o})\nn\\
\sum_{\pi^e}\prod_{ l=0\; {\rm even}}^{n-1}(-1)^{\sigma^e} x_{\pi^e(l)}^{2l+1} I_{\nu+1}(x_{\pi^e(l)})
&=&\Delta(x^2_{k^e}) \prod_{k^e \; even} I_{\nu+1}(x_{k^e}).
\ee
Next we combine the Vandermonde determinants as
\be
\frac {\Delta(x^2_{k^o}) \Delta(x^2_{k^e}) \Delta(x_k)}{\Delta(x_k^2)}&=&
\frac {\Delta(x_{k^e})\Delta(x_{k^o}) \Gamma(x_{k^o},x_{k^e})}
{ \Gamma(x_{k^o}^2,x_{k^e}^2)}\nn \\&=&
\frac {\Delta(x_{k^e})\Delta(x_{k^o}) }
{ \Gamma(x_{k^o},-x_{k^e})}\nn \\
&=& \det \frac 1{ x_{k^o}+x_{l^e}}
\ee
with
\be
\Gamma(x_k,y_k) = \prod_{k,l}(x_k-y_l).
\ee
The combination $D$ can thus be written as
\be
D =\sum_{\pi^{eo}}(-1)^{\sigma^{eo}} 
\det \frac {I_\nu(x_{k^o})x_{l^e}I_{\nu+1}(x_{l^e})} { x_{k^o}+x_{l^e}}.
\ee
The determinant is a sum over permutations of even
and odd integers which together with $\pi^{eo}$ can be combined into
a sum over all permutations
\be
D =\sum_{\pi}(-1)^{\sigma} 
\frac {I_\nu(x_{\pi(k)})x_{\pi(l)}I_{\nu+1}(x_{\pi(l)})} { x_{\pi(k)}+x_{\pi(l)}},
\ee
which is equal to the Pfaffian 
\be
D ={\rm Pf}\left [\frac{ I_\nu(x_{k})x_{l}I_{\nu+1}(x_{l})- I_\nu(x_{l})x_{k}I_{\nu+1}(x_{k})} { x_{k}+x_{l}}\right ],
\ee
where we have recovered the Pfaffian structure of \cite{K}. This 
leads to \cite{K}
\be
\label{ZNfPfaffian}
Z^\nu_{N_f}({\cal M};\ha_8) = \frac{1}{\Delta({\cal M})} 
{\rm Pf}[(\hm_j-\hm_i)Z^\nu_{N_f=2}(\hm_j,\hm_i;\ha_8)]_{j,i=1,\ldots,N_f}.
\ee
The alternative proof given 
here shows that the result is manifestly universal.

\subsection{The four flavor partition function}

For the four flavor partition function entering Eq. (\ref{rhocNf2}) 
the Pfaffian structure yields
\be
Z^\nu_{N_f=4}(\hz,\hz^*,\hm_3,\hm_4;\ha_8) & = & \frac{Z^\nu_2(\hz,\hz^*;\ha_8)Z^\nu_2(\hm_3,\hm_4;\ha_8)}{(\hz-\hm_3)(\hz-\hm_4)(\hz^*-\hm_3)(\hz^*-\hm_4)} \\
&&-\frac{Z^\nu_2(\hz,\hm_3;\ha_8)Z^\nu_2(\hz^*,\hm_4;\ha_8)}{(\hz-\hz^*)(\hz-\hm_4)(\hz^*-\hm_3)(\hm_3-\hm_4)} \nn \\
&&+ \frac{Z^\nu_2(\hz^*,\hm_3;\ha_8)Z^\nu_2(\hz,\hm_4;\ha_8)}{(\hz-\hz^*)(\hz-\hm_3)(\hz^*-\hm_4)(\hm_3-\hm_4)} \nn .
\ee
The latter two terms form a derivative in the limit $\hm_3\to \hm_4 = \hm$
\be
Z^\nu_4(\hz,\hz^*,\hm,\hm;\ha_8) & = & \frac{Z^\nu_2(\hz,\hz^*;\ha_8)Z^\nu_2(\hm,\hm;\ha_8)}{(\hz-\hm)^2(\hz^*-\hm)^2} \\
& & - \frac{\partial_{\hm}[\hat{Z}^\nu_2(\hz,\hm;\ha_8)]\hat{Z}^\nu_2(\hz^*,\hm;\ha_8)-\hat{Z}^\nu_2(\hz,\hm;\ha_8)\partial_{\hm}[\hat{Z}^\nu_2(\hz^*,\hm;\ha_8)]}{(\hz-\hz^*)(\hz-\hm)^2(\hz^*-\hm)^2}\nn.
\ee
With this we have succeeded in expressing the four flavor partition 
function in terms of the two flavor partition function. This form 
inserted in Eq. (\ref{rhocNf2}) leads to Eq. (\ref{rhocNf2fac}).

\section{Mean field including fluctuations}
\label{app:MF}

Here we compute the mean field eigenvalue density of $D_W$ including 
the fluctuations about the saddle points. 
In \ref{app:MF1} we derive the mean field limit of the two flavor 
partition function. A mean field approximation for the four flavor 
partition function that enters in the spectral density, 
\eqref{rhoc_micro_a6a8}, is given in \ref{app:MF2}, and the mean field 
result for the spectral density is derived in \ref{app:MF3}. We discuss 
the explicit dependence on the low energy constants $W_6$ and $W_8$ and 
give the result both for the Aoki phase and the Sharpe-Singleton scenario.
As explained in section \ref{sec:constraints} we have $W_6<0$ and $W_8>0$.

\subsection{The two flavor partition function}\label{app:MF1}

We consider the two-flavor partition function
\begin{eqnarray}\label{a1.1}
  Z_2^\nu(\hm;\ha_6,\ha_8)&=&\int\limits_{\U(2)}\exp\left[\frac{\hm}{2}\tr(U+U^{-1})+|\ha_6|^2[\tr(U+U^{-1})]^2-\ha_8^2\tr(U^2+U^{-2})\right]\\
 &\times&{\det}^\nu Ud\mu(U)\nonumber\\
 &=&\frac{1}{2\pi^2}\int\limits_{[0,2\pi]^2}\exp\left[\hm(\cos\varphi_1+\cos\varphi_2)+4|\ha_6|^2(\cos\varphi_1+\cos\varphi_2)^2\right]\nonumber\\
 &\times&\exp\left[-4\ha_8^2(\cos^2\varphi_1+\cos^2\varphi_2)+4\ha_8^2\right]e^{\imath\nu(\varphi_1+\varphi_2)}\sin^2\left(\frac{\varphi_1-\varphi_2}{2}\right)d[\varphi]\nonumber\\
 &=&\frac{1}{2\pi^2}\int\limits_{[0,2\pi]^2}\exp\left[-2(\ha_8^2-2|\ha_6|^2)\left(\cos\varphi_1+\cos\varphi_2-\frac{\hm}{4(\ha_8^2-2|\ha_6|^2)}\right)^2\right]e^{\imath\nu(\varphi_1+\varphi_2)}\nonumber\\
 &\times&\exp\left[-2\ha_8^2(\cos\varphi_1-\cos\varphi_2)^2+4\ha_8^2+\frac{\hm^2}{8(\ha_8^2-2|\ha_6|^2)}\right]\sin^2\left(\frac{\varphi_1-\varphi_2}{2}\right)d[\varphi].\nonumber
\end{eqnarray}
From the exponent we recognize that in the mean field limit we always have
\begin{eqnarray}
 \cos\varphi_1&=&\cos\varphi_2.\label{a1.2}
\end{eqnarray}

For $\ha_8^2+2\ha_6^2<0$ the solution of
\begin{eqnarray}
 \cos\varphi_1+\cos\varphi_2&=&\frac{\hm}{4(\ha_8^2-2|\ha_6|^2)}\label{a1.3}
\end{eqnarray}
is a minimum and does not contribute in the mean field limit (this is the 
case of the Sharpe-Singleton scenario). Therefore the maxima can only come from
\begin{eqnarray}
 \sin\varphi_1&=&\sin\varphi_2=0.\label{a1.4}
\end{eqnarray}
In combination with Eq.~\eqref{a1.2} this yields the two solutions $\cos\varphi_1=\cos\varphi_2=\pm1$.

We make the following expansion
\begin{eqnarray}
 \varphi_{1/2}^{(+)}&=&\delta\varphi_{1/2},\quad\cos\varphi_{1/2}^{(+)}=1-\frac{1}{2}\delta\varphi_{1/2}^2,\nonumber\\
 \varphi_{1/2}^{(-)}&=&\pi+\delta\varphi_{1/2},\quad\cos\varphi_{1/2}^{(-)}=-1+\frac{1}{2}\delta\varphi_{1/2}^2.\label{a1.5}
\end{eqnarray}
The maximum of the two points is at $\cos\varphi_{1/2}={\rm sign} \, \hm$. Thus we obtain the two flavor partition function
\begin{eqnarray}\label{a1.6}
  Z_2^{\rm MF}(\hm;\ha_6,\ha_8)&=&\frac{1}{8\pi^2}\exp\left[2|\hm|+16|\ha_6|^2-4\ha_8^2\right]\\
 &\times&\int\limits_{\mathbb{R}^2}\exp\left[-\left(\frac{|\hm|}{2}+8|\ha_6|^2-4\ha_8^2\right)(\delta\varphi_1^2+\delta\varphi_2^2)\right](\delta\varphi_1-\delta\varphi_2)^2d[\delta\varphi]\nonumber\\
 &=&\frac{\exp\left[2|\hm|+16|\ha_6|^2-4\ha_8^2\right]}{2\pi(|\hm|+16|\ha_6|^2-8\ha_8^2)^2}\nonumber
\end{eqnarray}
for $\ha_8^2+2\ha_6^2<0$.

For $\ha_8^2+2\ha_6^2>0$ (i.e.~in the Aoki phase) the saddlepoint given in Eq.~\eqref{a1.3}, is a maximum. Hence we have to take it into account in the saddlepoint analysis if the right hand side of Eq.~\eqref{a1.3} is in the interval $[-2,2]$. Thereby we recognize that there are actually four saddlepoints fulfilling both conditions~\eqref{a1.2} and \eqref{a1.3}. The two angles may have the same sign or the opposite one. Those with the same sign are algebraically suppressed by the $\sin^2$ in the measure.

Let $\varphi_0={\rm arcos}(\hm/(8\ha_8^2-16|\ha_6|^2))$. The expansion about $\pm\varphi_0$ is given by
\begin{eqnarray}
 \varphi_{1/2}^{(+)}&=&\pm\varphi_0+\delta\varphi_{1/2},\quad\cos\varphi_{1/2}^{(+)}=\frac{\hm}{8\ha_8^2-16|\ha_6|^2}\mp\sin\varphi_0\delta\varphi_{1/2},\nonumber\\
 \varphi_{1/2}^{(-)}&=&\mp\varphi_0+\delta\varphi_{1/2},\quad\cos\varphi_{1/2}^{(-)}=\frac{\hm}{8\ha_8^2-16|\ha_6|^2}\pm\sin\varphi_0\delta\varphi_{1/2}.\label{a1.7}
\end{eqnarray}
The simplified integral which we have to solve is
\begin{eqnarray}\label{a1.8}
  &&\frac{1}{2\pi^2}\int\limits_{\mathbb{R}^2}\exp\left[-2(\ha_8^2-2|\ha_6|^2)\sin^2\varphi_0\left(\delta\varphi_1-\delta\varphi_2\right)^2\right]e^{\imath\nu(\varphi_1+\varphi_2)}\\
 &\times&\exp\left[-2\ha_8^2\sin^2\varphi_0(\delta\varphi_1+\delta\varphi_2)^2+4\ha_8^2+\frac{\hm^2}{8(\ha_8^2-2|\ha_6|^2)}\right]\sin^2\varphi_0d[\delta\varphi]\nonumber\\
 &=&\frac{1}{8\pi\sqrt{\ha_8^2(\ha_8^2-2|\ha_6|^2)}}\exp\left[4\ha_8^2+\frac{\hm^2}{8(\ha_8^2-2|\ha_6|^2)}\right].\nonumber
\end{eqnarray}
Hence the two flavor partion function is given by
\begin{eqnarray}\label{a1.9}
  Z_2^{\rm MF}(\hm;\ha_6,\ha_8)&=&\frac{\exp\left[2|\hm|+16|\ha_6|^2-4\ha_8^2\right]}{2\pi(|\hm|+16|\ha_6|^2-8\ha_8^2)^2}\\
 &+&\frac{1}{4\pi\sqrt{\ha_8^2(\ha_8^2-2|\ha_6|^2)}}\exp\left[4\ha_8^2+\frac{\hm^2}{8(\ha_8^2-2|\ha_6|^2)}\right]\theta(8\ha_8^2-16|\ha_6|^2-|\hm|)\nonumber
\end{eqnarray}
for $\ha_8^2+2\ha_6^2>0$. Please notice that the second term results from two saddlepoints at $\pm\varphi_0$ and only appears in a certain range of the quark mass. Moreover the result~\eqref{a1.9} is also valid for $\ha_8^2+2\ha_6^2<0$ since the Heavyside distribution vanishes in this regime.

\subsection{The modified four flavor partition function}\label{app:MF2}

We consider the four flavor partition function which for $W_6<0$ can be written as 
\begin{eqnarray}
  &&\widetilde{Z}_4^\nu(\hz,\hz^*,\hm;\ha_6,\ha_8)=\frac{\sqrt{\ha_8^2+2|\ha_6|^2}}{4\sqrt{\pi}|\ha_6|\ha_8}|\hy||\hz-\hm|^4\int\limits_{\mathbb{R}} dy_6\exp\left[-\frac{y_6^2}{16|\ha_6|^2}-\frac{(\hx-y_6)^2}{8\ha_8^2}-4\ha_8^2\right]\nonumber\\
 &\times&Z_4^\nu(\hz-y_6,\hz^*-y_6,\hm-y_6;\ha_8)\label{a2.1}\\
 &=&|\hy||\hz-\hm|^4\int\limits_{\U(4)}\exp\left[\frac{1}{2}\tr\,\diag(\hm,\hm,\hz,\hz^*)(U+U^{-1})-\ha_8^2\tr(U^2+U^{-2})\right]\nonumber\\
 &\times&\exp\left[\frac{4|\ha_6|^2\ha_8^2}{\ha_8^2+2|\ha_6|^2}\left(\frac{1}{2}\Tr(U+U^{-1})-\frac{\hx}{4\ha_8^2}\right)^2-\frac{\hx^2}{8\ha_8^2}-4\ha_8^2\right]{\det}^\nu Ud\mu(U)\nonumber\\
 &=&\imath\frac{64}{\pi^4}{\rm sign}(\hy)\int\limits_{[0,2\pi]^4}d[\varphi]\prod\limits_{1\leq i<j\leq4}\sin^2\left(\frac{\varphi_i-\varphi_j}{2}\right)\exp\left[-4\ha_8^2\sum\limits_{j=1}^4\cos^2\varphi_j+8\ha_8^2\right]\nonumber\\
 &\times&\exp\left[\frac{4|\ha_6|^2\ha_8^2}{\ha_8^2+2|\ha_6|^2}\left(\sum\limits_{j=1}^4\cos\varphi_j-\frac{x}{4\ha_8^2}\right)^2-\frac{\hx^2}{8\ha_8^2}-4\ha_8^2+\imath\nu\sum\limits_{j=1}^4\varphi_j\right]\nonumber\\
 &\times&\frac{\det\left[\exp[\hm\cos\varphi_j],\cos\varphi_j\exp[\hm\cos\varphi_j],\exp[\hz\cos\varphi_j],\exp[\hz^*\cos\varphi_j]\right]}{\prod\limits_{1\leq i<j\leq4}(\cos\varphi_i-\cos\varphi_j)}\nonumber\\
 &=&\frac{32}{\pi^4}\exp[4\ha_8^2]\int\limits_{[0,2\pi]^4}d[\varphi]\prod\limits_{1\leq i<j\leq4}\sin^2\left(\frac{\varphi_i-\varphi_j}{2}\right)e^{\imath\nu(\varphi_1+\varphi_2+\varphi_3+\varphi_4)}\sum\limits_{\omega\in\mathbf{S}(4)}\nonumber\\
 &\times&\frac{\exp\left[-2\ha_8^2(\cos\varphi_{\omega(1)}-\cos\varphi_{\omega(2)})^2-2\ha_8^2(\cos\varphi_{\omega(3)}-\cos\varphi_{\omega(4)})^2\right]}{(\cos\varphi_{\omega(1)}-\cos\varphi_{\omega(3)})(\cos\varphi_{\omega(1)}-\cos\varphi_{\omega(4)})(\cos\varphi_{\omega(2)}-\cos\varphi_{\omega(3)})(\cos\varphi_{\omega(2)}-\cos\varphi_{\omega(4)})}\nonumber\\
 &\times&\frac{\sin\left[|\hy|(\cos\varphi_{\omega(3)}-\cos\varphi_{\omega(4)})\right]}{\cos\varphi_{\omega(3)}-\cos\varphi_{\omega(4)}}\nonumber\\
 &\times&\exp\left[(4|\ha_6|^2-2\ha_8^2)(\cos\varphi_{\omega(1)}+\cos\varphi_{\omega(2)})^2+\hm(\cos\varphi_{\omega(1)}+\cos\varphi_{\omega(2)})\right]\nonumber\\
 &\times&\exp\left[-\frac{1}{8(\ha_8^2+2|\ha_6|^2)}[\hx+8|\ha_6|^2(\cos\varphi_{\omega(1)}+\cos\varphi_{\omega(2)})-4\ha_8^2(\cos\varphi_{\omega(3)}+\cos\varphi_{\omega(4)})]^2\right].\nonumber
\end{eqnarray}
The permutation group of four elements is denoted by $\mathbf{S}(4)$.

In the mean field limit we have to expand the partition function about the maxima of the exponent. Omitting the permutations we identify two imediate conditions,
\begin{eqnarray}\label{a2.2}
 \cos\varphi_1^{(0)}=\cos\varphi_2^{(0)}\quad\mathrm{and}\quad\cos\varphi_3^{(0)}=\cos\varphi_4^{(0)}.
\end{eqnarray}
This is solved by
\begin{eqnarray}\label{a2.3}
 \varphi_1^{(0)}=-\varphi_2^{(0)}\quad\mathrm{and}\quad\varphi_3^{(0)}=-\varphi_4^{(0)}.
\end{eqnarray}
Other choices are supressed by the Vandermonde determinant. Hence we have to maximize the function
\begin{eqnarray}\label{a2.4}
  f(x,\varphi_1)&=&\exp\left[8(2|\ha_6|^2-\ha_8^2)\cos^2\varphi_{1}+2\hm\cos\varphi_{1}\right]\nonumber\\
 &\times&\exp\left[-\frac{1}{8(\ha_8^2+2|\ha_6|^2)}[\hx+16|\ha_6|^2\cos\varphi_{1}-8\ha_8^2\cos\varphi_{3}]^2\right].
\end{eqnarray}

We consider the case $\ha_8^2+2\ha_6^2<0$ (the Sharpe-Singleton scenario). Therefore the extremum for $\cos\varphi_{1}$ is a minimum and not a maximum. The situation would be completely different for $\ha_8^2+2\ha_6^2>0$, see discussion after Eq.~\eqref{a2.13}.

The maximum of $f(x,\varphi_1)$ for all $x$ is given by
\begin{eqnarray}\label{a2.5}
 \underset{x\in\mathbb{R}}{\max}f(x,\varphi_1)=\exp\left[8(2|\ha_6|^2-\ha_8^2)\cos^2\varphi_{1}+2\hm\cos\varphi_{1}\right].
\end{eqnarray}
This result takes its maximum at $\cos\varphi_{1}^{(0)}={\rm sign}\, m$ yielding
\begin{eqnarray}\label{a2.6}
 \underset{x\in\mathbb{R},\varphi_1\in[0,2\pi]}{\max}f(x,\varphi_1)=\exp\left[16|\ha_6|^2-8\ha_8^2+2|\hm|\right].
\end{eqnarray}
In the integral~\eqref{a2.1} this maximum should be inside the interval
\begin{eqnarray}\label{a2.7}
 \hx\in\left[-8\ha_8^2-16|\ha_6|^2{\rm sign}\, \hm ,8\ha_8^2-16|\ha_6|^2{\rm sign}\, \hm\right].
\end{eqnarray}
The condition for the second integral is then
\begin{eqnarray}\label{a2.8}
 \cos\varphi_3^{(0)}=\frac{\hx+16|\ha_6|^2{\rm sign}\, \hm}{8\ha_8^2}.
\end{eqnarray}
We make the following expansion
\begin{eqnarray}\label{a2.9}
 \varphi_1&=&\frac{1-{\rm sign}\, \hm}{2}\pi+\delta\varphi_1,\quad\cos\varphi_1={\rm sign}\, \hm-\frac{{\rm sign}\, \hm}{2}\delta\varphi_1^2,\\
 \varphi_2&=&-\frac{1-{\rm sign}\, \hm}{2}\pi+\delta\varphi_2,\quad\cos\varphi_2={\rm sign}\, \hm-\frac{{\rm sign}\, \hm}{2}\delta\varphi_2^2,\nonumber\\
 \varphi_3&=&\varphi_3^{(0)}+\delta\varphi_3,\quad\cos\varphi_3=\cos\varphi_3^{(0)}-\sin\varphi_3^{(0)}\delta\varphi_3,\nonumber\\
 \varphi_4&=&-\varphi_3^{(0)}+\delta\varphi_4,\quad\cos\varphi_4=\cos\varphi_3^{(0)}+\sin\varphi_3^{(0)}\delta\varphi_4.\nonumber
\end{eqnarray}
This expansion is substituted into Eq.~\eqref{a2.1} and we omit the sum since each term gives the same contribution and the degeneracy of the maximum,
\begin{eqnarray}
  &&\widetilde{Z}_4^{\rm MF}(\hz,\hz^*,\hm;\ha_6,\ha_8)=24\left(\frac{2}{\pi}\right)^4\exp[4\ha_8^2]\int\limits_{\mathbb{R}^4}d[\delta\varphi]\sin^2\varphi_3^{(0)}\sin^8\left(\frac{1-{\rm sign}\, \hm}{4}\pi-\frac{\varphi_3^{(0)}}{2}\right)\nonumber\\
 &\times&\left(\delta\varphi_1-\delta\varphi_2\right)^2\frac{\exp\left[-2\ha_8^2\sin^2\varphi_3^{(0)}(\delta\varphi_{3}+\delta\varphi_{4})^2\right]}{({\rm sign}\, \hm-(\hx+16|\ha_6|^2{\rm sign}\, \hm)/8\ha_8^2)^4}\frac{\sin\left[|\hy|\sin\varphi_3^{(0)}(\delta\varphi_{3}+\delta\varphi_{4})\right]}{\sin\varphi_3^{(0)}(\delta\varphi_{3}+\delta\varphi_{4})}\label{a2.10}\\
 &\times&\exp\left[-\left(\frac{|\hm|}{2}+8|\ha_6|^2-4\ha_8^2\right)(\delta\varphi_{1}^2+\delta\varphi_{2}^2)+2|\hm|+16|\ha_6|^2-8\ha_8^2\right]\nonumber\\
 &\times&\exp\left[-\frac{2\ha_8^4}{(\ha_8^2+2|\ha_6|^2)}\sin^2\varphi_3^{(0)}[\delta\varphi_{3}-\delta\varphi_{4}]^2\right]\theta(8\ha_8^2-|\hx+16|\ha_6|^2{\rm sign}\, \hm|).\nonumber
\end{eqnarray}
This integral decouples into two two-fold integrals. We need the following integral for large $|y|$,
\begin{eqnarray}\label{a2.11}
 \int\limits_{\mathbb{R}}\exp[-2\ha_8^2\lambda^2]\frac{\sin(|\hy|\lambda)}{\lambda}d\lambda&=&\pi{\rm erf}\left[\frac{|\hy|}{\sqrt{8}\ha_8}\right]\overset{|\hy|\gg1}{=}\pi,
\end{eqnarray}
where ${\rm erf}$ is the error function and use the identity
\begin{eqnarray}
 \sin^8\left(\frac{1-{\rm sign}\, \hm}{4}\pi-\frac{\varphi_3^{(0)}}{2}\right)&=&\frac{1}{2}\left(1-\cos\left(\frac{1-{\rm sign}\, \hm}{2}\pi-\varphi_3^{(0)}\right)\right)^4\nonumber\\
 &=&\frac{1}{16}\left(1-{\rm sign}\, \hm\cos\varphi_3^{(0)}\right)^4\nonumber\\
 &=&\frac{1}{16}({\rm sign}\, \hm-(\hx+16|\ha_6|^2{\rm sign}\, \hm)/8\ha_8^2)^4.\label{a2.12}
\end{eqnarray}
Then the final result for the partition function is given by
\begin{eqnarray}
  \widetilde{Z}_4^{\rm MF}(\hz,\hz^*,\hm;\ha_6,\ha_8)&=&3\left(\frac{2}{\pi}\right)^{3/2}\frac{\sqrt{\ha_8^2+2|\ha_6|^2}}{\ha_8^2}\frac{\exp\left[2|\hm|+16|\ha_6|^2-4\ha_8^2\right]}{(|\hm|/2+8|\ha_6|^2-4\ha_8^2)^2}\nonumber\\
 &\times&\theta(8\ha_8^2-|\hx+16|\ha_6|^2{\rm sign}\, \hm|)\label{a2.13}
\end{eqnarray}
for $\ha_8^2+2\ha_6^2<0$ (in the Sharpe Singleton scenario).

In the Aoki phase, $\ha_8^2+2\ha_6^2>0$, the extremum for $\cos\varphi_1$ is a maximum, cf. Eq.~\eqref{a2.4}. However it will only contribute if
\begin{eqnarray}\label{a2.14}
 |\hm|\leq8\ha_8^2-16|\ha_6|^2
\end{eqnarray}
and
\begin{eqnarray}\label{a2.15}
 \left|\hx+\frac{2|\ha_6|^2\hm}{\ha_8^2-2|\ha_6|^2}\right|\leq8\ha_8^2.
\end{eqnarray}
Then the saddlepoint changes to
\begin{eqnarray}\label{a2.16}
 \cos\varphi_1^{(0)}&=&\frac{\hm}{8\ha_8^2-16|\ha_6|^2},\\
 \cos\varphi_3^{(0)}&=&\frac{\hx}{8\ha_8^2}+\frac{|\ha_6|^2\hm}{4\ha_8^2(\ha_8^2-2|\ha_6|^2)}.\nn
\end{eqnarray}
Hence the expansion about the saddle points is given by
\begin{eqnarray}\label{a2.17}
 \varphi_1&=&\varphi_1^{(0)}+\delta\varphi_1,\quad\cos\varphi_1=\cos\varphi_1^{(0)}-\sin\varphi_1^{(0)}\delta\varphi_2,\\
 \varphi_2&=&-\varphi_1^{(0)}+\delta\varphi_2,\quad\cos\varphi_2=\cos\varphi_1^{(0)}+\sin\varphi_1^{(0)}\delta\varphi_2,\nonumber\\
 \varphi_3&=&\varphi_3^{(0)}+\delta\varphi_3,\quad\cos\varphi_3=\cos\varphi_3^{(0)}-\sin\varphi_3^{(0)}\delta\varphi_3,\nn\\
 \varphi_4&=&-\varphi_3^{(0)}+\delta\varphi_4,\quad\cos\varphi_4=\cos\varphi_3^{(0)}+\sin\varphi_3^{(0)}\delta\varphi_4.\nonumber
\end{eqnarray}
The degeneracy of this maximum is four which results in the integral
\begin{eqnarray}
 &&3\frac{2^{10}}{\pi^4}\exp[4\ha_8^2]\int\limits_{\mathbb{R}^4}d[\delta\varphi]\sin^2\varphi_1^{(0)}\sin^2\varphi_3^{(0)}\sin^4\left(\frac{\varphi_1^{(0)}-\varphi_3^{(0)}}{2}\right)\sin^4\left(\frac{\varphi_1^{(0)}+\varphi_3^{(0)}}{2}\right)\nonumber\\
 &\times&\frac{\exp\left[-2\ha_8^2\sin^2\varphi_1^{(0)}(\delta\varphi_{1}+\delta\varphi_{2})^2-2\ha_8^2\sin^2\varphi_3^{(0)}(\delta\varphi_{3}+\delta\varphi_{4})^2\right]}{(\cos\varphi_{1}^{(0)}-\cos\varphi_{3}^{(0)})^4}\nonumber\\
 &\times&\frac{\sin\left[|\hy|\sin\varphi_3^{(0)}(\delta\varphi_{3}+\delta\varphi_{4})\right]}{\sin\varphi_3^{(0)}(\delta\varphi_{3}+\delta\varphi_{4})}\nonumber\\
 &\times&\exp\left[(4|\ha_6|^2-2\ha_8^2)\sin^2\varphi_1^{(0)}(\delta\varphi_{1}+\delta\varphi_{2})^2+\frac{\hm^2}{8\ha_8^2-16|\ha_6|^2}\right]\nonumber\\
 &\times&\exp\left[-\frac{2}{\ha_8^2+2|\ha_6|^2}\left[2|\ha_6|^2\sin\varphi_1^{(0)}(\delta\varphi_{1}-\delta\varphi_{2})-\ha_8^2\sin\varphi_3^{(0)}(\delta\varphi_{3}-\delta\varphi_{4})\right]^2\right]\nonumber\\
 &=&6\left(\frac{2}{\pi}\right)^{3/2}\frac{1}{\ha_8^2}\sqrt{\frac{\ha_8^2+2|\ha_6|^2}{\ha_8^2-2|\ha_6|^2}}\exp\left[\frac{\hm^2}{8\ha_8^2-16|\ha_6|^2}+4\ha_8^2\right].\label{a2.18}
\end{eqnarray}
Combining this with the result~\eqref{a2.13} for $\ha_8^2+2\ha_6^2<0$ we find
\begin{eqnarray}
  &&\widetilde{Z}_4^{\rm MF}(\hz,\hz^*,\hm;\ha_6,\ha_8)=3\left(\frac{2}{\pi}\right)^{3/2}\frac{\sqrt{\ha_8^2+2|\ha_6|^2}}{\ha_8^2}\frac{\exp\left[2|\hm|+16|\ha_6|^2-4\ha_8^2\right]}{(|\hm|/2+8|\ha_6|^2-4\ha_8^2)^2}\nonumber\\
 &\times&\theta(8\ha_8^2-|\hx+16|\ha_6|^2{\rm sign}\, \hm|)+6\left(\frac{2}{\pi}\right)^{3/2}\frac{1}{\ha_8^3}\sqrt{\frac{\ha_8^2+2|\ha_6|^2}{\ha_8^2-2|\ha_6|^2}}\nonumber\\
 &\times&\exp\left[\frac{\hm^2}{8\ha_8^2-16|\ha_6|^2}+4\ha_8^2\right]\theta(8\ha_8^2-16|\ha_6|^2-|\hm|)\theta\left(8\ha_8^2-\left|\hx+\frac{2|\ha_6|^2\hm}{(\ha_8^2-2|\ha_6|^2)}\right|\right)\label{a2.19}.
\end{eqnarray}
This formula applies to both scenarios since the Heavyside distribution puts the second term to zero in the Sharpe-Singleton scenario.

\subsection{The unquenched level density}\label{app:MF3}

Combining the mean field limit of the numerator and denominator of Eq.~\eqref{rhoc_micro_a6a8} given by Eq.~\eqref{a1.9} and Eq.~\eqref{a2.19}, respectively, we obtain
\begin{eqnarray}\label{a3.1}
 \rho_{c,N_f=2}^{\rm MF}(\hx,\hm;\ha_6,\ha_8)&=&\frac{1}{32(2\pi)^{5/2}\sqrt{\ha_8^2+2|\ha_6|^2}}\frac{\widetilde{Z}_4^{\rm MF}(\hz,\hz^*,\hm;\ha_6,\ha_8)}{Z_2^{\rm MF}(\hm;\ha_6,\ha_8)}\\
 &=&\frac{3}{(2\pi)^3}\frac{1}{\ha_8^2}\left[\theta(2|\ha_6|^2-\ha_8^2)\theta(8\ha_8^2-|\hx+16|\ha_6|^2{\rm sign}\, \hm|)\right.\nonumber\\
&+&\left.\theta(8\ha_8^2-16|\ha_6|^2-|\hm|)\theta\left(8\ha_8^2-\left|\hx+\frac{2|\ha_6|^2\hm}{(\ha_8^2-2|\ha_6|^2)}\right|\right)\right]\nonumber
\end{eqnarray}
independent of the value of $W_6$. The first term will drop out if we are in the Aoki phase whereas the second term vanishes in the Sharpe-Singleton scenario. However the reason for this mechanism is quite different in the two cases. In the Aoki phase the first term is exponentially supressed in comparison to the second one which results from the extremum~\eqref{a2.16}. In the Sharpe-Singleton scenario the saddlepoint is a minimum and enters \textit{a priori} not in the saddlepoint analysis. Hence we have to look at the boundaries of the four dimensional box spanned by the four cosinus, see the discussion in \ref{app:MF2}.

This mechanism explains why we find a second order transition in the Aoki phase and a first order transition in the Sharpe-Singleton scenario. The extremum~\eqref{a2.16} can cross the four dimensional box with varying quark mass $\hm$ and eigenvalue $\hx$. Hence we have a continuous process from one boundary to the other in the Aoki scenario. When this extremum is excluded as in the Sharpe-Singleton scenario, the maximum has to jump from one boundary to the other. This manifests itself in the sign of the mass in the Heavyside distribution of the first term and the mass itself in the other one.



\end{document}